\documentclass[11pt]{article}
\usepackage{amssymb,amsmath,graphicx}
\usepackage{amsfonts}
\usepackage[caption=false]{subfig}
\usepackage{epsfig,latexsym,graphicx}
\usepackage[sort,numbers,compress]{natbib}
\usepackage{setspace,color}
\usepackage[normalem]{ulem}
\def\trd#1{\textcolor{red}{}}%

\begin{document}

\title{Robust Inference under the Beta Regression Model \\with Application to Health Care Studies}

\author{
Abhik Ghosh \\
Indian Statistical Institute, Kolkata, India \\
{\it abhianik@gmail.com}
}
\date{}
\maketitle


\begin{abstract}
Data on rates, percentages or proportions arise frequently in many {different} applied disciplines  
like medical biology, health care, psychology and several others.
In this paper, we develop a robust inference procedure for the beta regression model
which is used to describe such response variables taking values in $(0, 1)$ through some related explanatory variables.
{In relation to the beta regression model, the issue of robustness has been largely 
ignored in the literature so far.}
The existing maximum likelihood  based inference has serious lack of robustness against outliers in data
and generate drastically different (erroneous) inference in presence of data contamination.
Here, we develop the robust minimum density power divergence estimator
and a class of robust Wald-type tests for the beta regression model along with several applications. 
We derive their asymptotic properties and describe their robustness theoretically through the influence function analyses. 
Finite sample performances of the proposed estimators and tests are examined through suitable simulation studies
and real data applications in the context of health care and psychology. 
{Although we primarily focus on the beta regression models with a fixed dispersion parameter, 
	some indications are also provided for extension to the variable dispersion beta regression models
	with an application.}
\end{abstract}



\section{Introduction}\label{SEC:intro}

In many biological experiments, medical {research} including health care studies
and psychology, survey {research} in sociology and marketing, and several other applied sciences, 
we often come across data on rates, ratios, percentages or proportions, taking values in the unit interval $(0, 1)$. 
Examples of such data include the ``body fat percentage" or any similar health condition measured in percentage,
health assessment questionnaire (HAQ) data or similar rating data, 
accuracy percentage of {any treatment in clinical trials}, 
experimental scores measuring stress, depression, etc.~in psychology, 
proportion of a certain group of patients (for some particular disease) in a region 
and many more. Such data can be modeled individually by a beta distribution having support $(0, 1)$.
However, in order to better understand  the underlying data-generating mechanism 
{for} more detailed inference, it is often required to relate {their} values with some other associated explanatory variables
through a suitable regression structure{; this} also {enables} us to do prediction. 
The beta regression model {(BRM)} is designed to help in this situation,
which {models} a response variable $y$ taking values in $(0, 1)$ through 
{a} set of explanatory variables $\boldsymbol{x}$.

There are several recent specifications of the {BRM};
for example, see \cite{Paolino:2001,Kieschnick/McCullough:2003,Ferrari/Cribari-Neto:2004,Vasconcellos/Cribari-Neto:2004}, 
among others.
In this paper, we follow the most popular specification provided by \cite{Ferrari/Cribari-Neto:2004}.
This is because {this specification} (i)   models the ``mean" of {the} response variable on $(0, 1)$ 
to depend on a linear combination of available covariates through a suitable link function, 
(ii)  is closely related to the popular class of generalized linear models 
\citep{McCullagh/Nelder:1989}, 
(iii)  allows many different possible link functions to model various structures within the data, 
and (iv) the inference methodologies are well developed for this specification
and are available in {the} standard statistical software R (package `{\it betareg}') for practitioners. 

Mathematically, suppose $y_1,\dots ,~y_n$ are $n$ independent responses each taking value in $(0, 1)$
and are associated with $p$-dimensional covariate values $\boldsymbol{x}_1, \ldots, \boldsymbol{x}_n$, respectively. 
Then, in the beta regression model (BRM) of \cite{Ferrari/Cribari-Neto:2004},
each $y_i$ follows a beta distribution having density {$f\left(\cdot; \mu_i,~{\phi }\right)$, where}
\begin{eqnarray}
f\left(y; \mu,~{\phi }\right)= 
\frac{{1}}{{B}\left(\mu{\phi },~\left(1-\mu\right){\phi }\right)}
~{y}^{\mu{\phi}-1}{\left({1}{-}{{y}}\right)}^{\left(1-\mu\right){\phi }-1},~~~~~~~~0<y<1,
\label{EQ:BRM_density}
\end{eqnarray}	 
{with $B(\cdot, \cdot)$ being the (complete) beta function,
and $E\left(y_i\right)={\mu}_i\in (0, 1)$ is related to the (given) {$i$-th} value 
{$\boldsymbol{x}_i \in \mathbb{R}^p$} of 
the explanatory variables through a suitable link function $g$ (defined on $(0, 1)$).
Note that, $Var\left(y_i\right)=\frac{{\mu }_i\left(1-{\mu }_i\right)}{1+\phi }$.
Given $g$, the BRM of \cite{Ferrari/Cribari-Neto:2004} --
with fixed precision parameter $\phi$ or dispersion parameter $\sigma^2 = (1+\phi)^{-1}$ -- 
assumes the regression structure}
\begin{eqnarray}
y_i \sim f\left(y_i; \mu_i, {\phi }\right) \mbox{ independently, with } 
{g}\left({{\mu }}_{{i}}\right){=}{\boldsymbol{x}}^{{T}}_{{i}}\boldsymbol{\beta },
~~i=1, \ldots, n,
\label{EQ:BRM}
\end{eqnarray}	 
where $\boldsymbol\beta\in \mathbb{R}^p$ is the vector of unknown regression coefficients.
Our objective then is to make inference about the parameter of interest 
$\boldsymbol\theta =(\boldsymbol\beta^T, \phi )^T\in\mathbb{R}^{p+1}$ based on the available data 
$\left\{\left(y_i, \boldsymbol{x}_i\right) : i=1, \ldots, n\right\}$.
{This BRM has later been extended to cover the cases of heterogeneous precision parameter $\phi_i$ 
	(or, dispersion parameter $\sigma_i^2$) for $y_i$ by 
	\cite{Smithson/Verkuilen:2006,Simas/etc:2010,Rocha/Simas:2011,Cribari-Neto/Souza:2012,Melo/etc:2015},
	where $\phi_i$ depends on another set of covariates through a (possibly different) regression structure.
	To keep a clear focus in our presentations, we restrict our attention primarily to the fixed dispersion 
	BRM (\ref{EQ:BRM}) in the present paper. 
	Our methodology, however, is not critically dependent on the fixed dispersion assumption,
	and we also briefly indicate the possible extension 
to a general class of non-linear, variable dispersion BRMs towards the end of the paper.
Indeed, our illustrations will show that the extension of the proposed methodology 
to such a general class of BRMs has exactly the same structure and robustness implications
in relation to the fixed dispersion results presented in this paper. }

{The BRM (\ref{EQ:BRM})} has become very useful in {several} recent applications,
since it can  {also} be applied to  data within any finite {interval}. 
If $y$ takes values in any other open interval, say $(a, b)$, 
we can apply the BRM  (\ref{EQ:BRM})  with the transformed response $\frac{y-a}{b-a}${$\in(0, 1)$}
\trd{which now takes values in $(0, 1)$}. 
Further, if the response $y$ also takes \trd{the} {values} in the end-points 0 and 1, 
{rather than using} sophisticated and complicated modifications, 
we can {simply} apply the \trd{simpler} BRM (\ref{EQ:BRM}) \trd{easily}
with the widely used ad-hoc transformation $\frac{1}{n}[y(n-1)+0.5]$, 
$n$ being the sample size \citep{Smithson/Verkuilen:2006}. 

The existing inference procedures under the BRM (\ref{EQ:BRM}) are primarily based on the classical maximum likelihood approach.
The point estimator  of $\boldsymbol\theta =(\boldsymbol\beta^T, \phi )^T$ is obtained by maximizing the 
likelihood function with respect to $\boldsymbol{\theta}$, generating the maximum likelihood estimator (MLE),
and any hypothesis testing problem can be solved by the likelihood ratio test or the Wald test based on the MLE;
see \cite{Ferrari/Cribari-Neto:2004} for more details. The R package `{\it betareg}' \trd{also} provides the inferential solution 
for the BRM (\ref{EQ:BRM}) based on this standard maximum likelihood approach,  
which {possesses} many asymptotic optimality properties. 
However, a serious problem with \trd{the} maximum likelihood based inference is the high degree of sensitivity
to potential outliers in the data. This lack of robustness often leads to drastically different (erroneous) 
inference in {the} presence of even a small amount of data contamination.
\trd{This non-robustness issue naturally affects the results also for the BRM 
while using the existing inference methodologies based on the likelihood principle in the presence of outliers.}
Since such outliers are not uncommon in practical datasets, 
we need to be very cautious before using \trd{the} maximum likelihood based inference
(and also using the R package `{\it betareg}').
To illustrate this issue, let us present a motivating example from an Australian health care study.

\bigskip
\noindent
\textbf{A Motivating Example (AIS Data):}\\
Consider the data on health measurements of several athletes 
collected at the Australian Institute of Sport (AIS) 
which is publicly available in the R package ``{\it sn}". 
{\citet{Bayes/etc:2012}} have recently studied {a subset of these data} corresponding to the 37 rowing athletes 
\trd{and tried} to predict their body fat percentages (BFP) from their lean body masses (LBM) using Bayesian inference.
Since the {response} variable BFP takes values within $(0, 1)$, 
{we here fit} the BRM (\ref{EQ:BRM}) with response $y=\mbox{BFP}$, covariate $\boldsymbol{x}_i=(1, \mbox{LBM})^T$
and a logit link function, $logit(E[\mbox{BFP}]) = \beta_1 + \beta_2 LBM$.
Then, applying the existing maximum likelihood procedure using `{\it betareg}', the MLE of the parameter of interest 
$\boldsymbol{\theta} = (\beta_1, \beta_2, \phi)^T$ {turns out to be} $(0.097, -0.027, 95.472)^T$.
Further, applying the existing Wald test based on this MLE, 
the p-values of the significance of two regression coefficients {become}
$0.699$ and $0$ respectively, which indicates that the intercept component ($\beta_1$) is not significant in the model.

\begin{figure}[!h]
	\centering
	\includegraphics[width=0.4\linewidth]{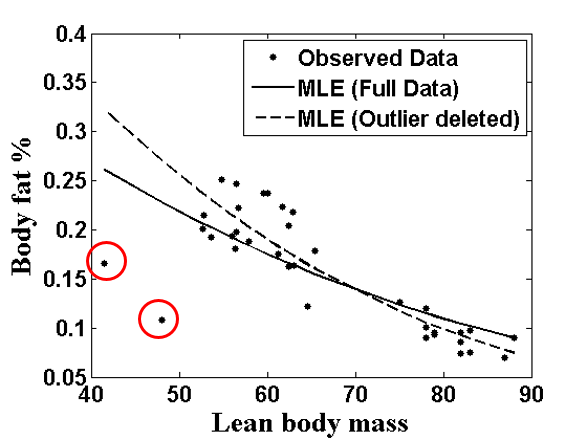}
	\caption{The AIS Data along with the fitted lines based on the MLE for the full data and the outlier deleted data.
		(The two outlying observations are marked with red circles).}
	\label{FIG:aismle}
\end{figure}

However, by plotting the data (see Figure \ref{FIG:aismle}), one can clearly see that 
there are two outlying observations as also noticed by \cite{Bayes/etc:2012}; 
the fitted line based on the above MLE does not yield a good fit to the bulk of the data in the presence of these two outliers.
In fact, if we again compute the MLE of the parameter $\boldsymbol{\theta}$ after removing these two outliers,
the resulting estimate becomes $(0.838, -0.038, 246.305)^T$, 
which drastically differs from the previous MLE based on the full data.
The change in the fitted line is clearly visible in Figure \ref{FIG:aismle}
and the estimate of $\phi$ changes substantially!
Further, after deleting these two outliers, both the p-values of the MLE based Wald test 
for testing the significance of the regression coefficients become $0$.
Thus, only these two outliers completely hide the significance of $\beta_1$ 
{reversing the conclusions of the} inference. 
\hfill{$\square$}

As we have seen in the above example, few outliers in a dataset can lead to completely wrong inference 
through the existing likelihood procedures under the BRM.
Several other authors have also recently noticed this non-robust behavior of the MLE 
and the instability of the related inferences against the outlying observations in the BRMs \citep{Ferrari/Cribari-Neto:2004,Espinheira/etc:2008a,Espinheira/etc:2008b};
they have developed some diagnostic tools to identify such influential observations or outliers in a BRM
and suggested {their deletion} before doing maximum likelihood based inference.
Although this solution with \trd{the} prior outlier detection is feasible 
{(although rarely advisable)} for \trd{the} simple and small datasets, 
it is quite difficult and needs several additional analyses for most complicated datasets
including the big or high-dimensional datasets of recent era.
A robust inference procedure that can automatically take care of these outliers to
successfully yield stable results {is} much more logical, efficient and useful in all such complicated cases. 
However, unlike other inferential set-ups, there exists no such robust inference procedure for 
the recently developed beta regression model. 
The only related work is the one by \cite{Bayes/etc:2012} 
who have proposed to solve this issue for the BRM under Bayesian paradigm
through the use of a modified distribution in place of the simple beta distribution;
but it does not really address the non-robustness problem of the MLE based inference 
with respect to the simpler specified {BRM}.

In this paper, we develop a robust inference procedure for the BRM (\ref{EQ:BRM})
without changing its original distributional form. 
Among several approaches of robust inference, we follow the minimum divergence approach
where we quantify the discrepancy between {the} data and the parametric model through a statistical divergence measure
and minimize it to estimate the unknown parameters.
In particular, we consider the density power divergence (DPD) of \cite{Basu/etc:1998}, because
the resulting estimator has become very popular in recent times due to its high asymptotic efficiency 
along with strong robustness properties. It has also been applied to many real life inference problems;
see Section \ref{SEC:Back_DPD} and \cite{Basu/etc:2011} for additional details. 
We develop the robust minimum density power divergence estimator (MDPDE)
for the BRM (\ref{EQ:BRM}) along with its asymptotic properties in Section \ref{SEC:Est}. 
Based on the proposed MDPDE, we develop a robust Wald-type hypothesis testing procedure 
in Section \ref{SEC:test} and derive its asymptotic properties. 
We also theoretically illustrate the robustness of both the proposed estimator and the testing procedure 
through suitable influence function analyses. 
{Finite sample performances of the proposed inference procedures are examined 
through suitable simulation studies in Section \ref{SEC:Simulation}. 
In Section \ref{SEC:Data}, our proposals are applied to reanalyze the motivating example 
along with two additional real data examples from health-care studies (including psychology).}
{Extension of the proposed methodology for performing robust inference under 
	a general class of (non--linear) variable dispersion BRM is briefly discussed in Section \ref{SEC:VarBRM}
and illustrated through a real data application.}
Finally{,} the paper ends with some concluding remarks in Section {\ref{SEC:concluding}}.

\section{Robust Minimum Density Power Divergence Estimators}
\label{SEC:Est}
 
\subsection{Background}\label{SEC:Back_DPD}

The density power divergence (DPD) measure between two densities $f_1$ and $f_2$
(with respect to some common dominating measure)
is defined in terms of a tuning parameter $\alpha\geq 0$ \citep{Basu/etc:1998} as
\begin{eqnarray}
{{d}}_{{\alpha }}\left(f_1, f_2 \right) &=& ~\int f_2^{1+\alpha} 
- \frac{1+\alpha}{\alpha} \int f_1 f_2^{\alpha} + \frac{1}{\alpha} \int f_1^{1+\alpha},
~~~~~~\mbox{if  }\alpha >0;
\label{EQ:DPD}\\
{{d}}_{{0}}\left(f_1, f_2 \right) &=& ~\lim_{\alpha \to 0} d_{\alpha} \left(f_1, f_2 \right)
= \int f_1 \log\left(\frac{f_1}{f_2}\right). \nonumber
\end{eqnarray} 
Note that the DPD measure at $\alpha=0$ coincides with the famous likelihood disparity,
minimization of which is known to {generate the MLE}.
The DPD family connects the likelihood disparity (at ${\alpha}$ = 0) to the $L_2$-Divergence (at ${\alpha}$ = 1) 
smoothly through the tuning parameter $\alpha$. For the sake of completeness and a better understanding,
let us start by recalling the minimum DPD estimation under the independent and identically distributed (iid) set-up.
	
For $n$ iid observations $Y_1,\ldots, Y_n$ modeled by a parametric family {of densities} 
$\mathcal{F}_{\boldsymbol{\theta}}=\left\{f_{\boldsymbol{\theta}}~:~\boldsymbol{\theta} \in \mathit{\Theta}\subseteq\mathbb{R}^k\right\}$,  
the {MDPDE} is obtained by minimizing the estimated DPD measure (\ref{EQ:DPD})
between the observed data (at $f_1$) and the model density $f_{\boldsymbol{\theta}}$ (at $f_2$),
or, equivalently by minimizing the quantity
\[
\int f^{1+\alpha}_{\boldsymbol{\theta}} - \frac{1+\alpha}{\alpha}\int f^{\alpha}_{\boldsymbol{\theta}}dG_n
=\int f^{1+\alpha}_{\boldsymbol{\theta}} - \frac{1+\alpha}{\alpha}\frac{1}{n} \sum^n_{i=1} f^{\alpha}_{\boldsymbol{\theta}}\left(Y_i\right),
\] 
with $G_n$ being the empirical distribution function based on the observed data \citep{Basu/etc:1998,Basu/etc:2011}. 
Under suitable differentiability assumptions, the estimating equation of $\boldsymbol{\theta}$ is given by 	
$$
\frac{1}{n} \sum^{n}_{i=1} \boldsymbol{u}_{\boldsymbol{\theta }}\left(Y_i\right) f^{\alpha}_{\boldsymbol{\theta}}\left(Y_i\right)
- \int \boldsymbol{u}_{\boldsymbol{\theta}}f^{1+\alpha}_{\boldsymbol{\theta}} = 0,
$$
{where $\boldsymbol{u}_{\boldsymbol{\theta}}=\nabla\ln f_{\boldsymbol{\theta}}$
	is the score function with $\nabla$ representing gradient with respect to $\boldsymbol{\theta}$.}
Note that, at ${\alpha}=0$, this MDPDE estimating equation coincides with the estimating (score) equation of the {MLE}.
The MDPDE at $\alpha>0$ yields a generalization of the MLE which down-weights the effect of the outlying observations
in the estimating equation by $\alpha$-th power of the model density and hence is expected to be more robust.
This MDPDE has become very popular in recent days, because
(i) it does not need non-parametric kernel estimation unlike many other divergences, 
(ii) it {is a} robust estimator having high asymptotic efficiency at properly chosen $\alpha$,
and (ii) it can be obtained from a simple unbiased estimating equation along with an underlying objective function
which helps to avoid the problem of multiple roots.

However, in general, our data for the BRM (\ref{EQ:BRM}) are NOT iid,
and hence the above approach cannot be applied directly. 
This is because we generally do not make any distributional assumptions on the covariates $\boldsymbol{x}_i$s 
and treat them as fixed (given) so that, for each $i$, 
$y_i\sim Beta\left({\mu }_i\phi ,~\left(1-{\mu }_i\right)\phi \right)$
with density given by (\ref{EQ:BRM}). 
Thus, each $y_i$ is independent but not identically distributed.
Recently, \cite{Ghosh/Basu:2013} have proposed an extension of the MDPDE 
for the general independent but non-homogeneous set-up by considering the average DPD measure over different distributions.
\cite{Ghosh/Basu:2016} {have} applied this extended approach to develop robust inference
for a simple class of canonical generalized linear models (GLMs) with fixed designs 
including normal, Poisson and logistic regressions;
\cite{Ghosh:2016} has also applied it to an exponential regression model to propose a robust estimator of the tail index. 
However, unfortunately, the class of GLMs considered in \cite{Ghosh/Basu:2016} does not 
directly cover our BRM (\ref{EQ:BRM}). 
So, in this paper, we further extend this approach to develop 
a robust estimator for the BRM (\ref{EQ:BRM}) with fixed covariates (design).
{For the sake of completeness, necessary background concepts, assumptions and results from \cite{Ghosh/Basu:2013}
	are presented in the online supplement.}

\subsection{The MDPDE for the Beta regression Model}\label{SEC:MDPDE_beta}

Consider the {BRM} set-up as {described} in Section \ref{SEC:intro}.
Let us assume that the responses $y_1, \dots, y_n$ are independent but $y_i~\sim g_i$ for each  $i=1, \dots, n$,
where $g_i$s are potentially different true densities of $y_i$s depending on $\boldsymbol{x}_i$s.
We model $g_i$ by the BRM given by (\ref{EQ:BRM}), i.e., by the model density
$ f_i\left(\cdot, \boldsymbol{\theta}\right) \equiv Beta\left(\mu_i{\phi }, \left(1-\mu_i\right){\phi}\right)$ density.
The unknown parameter of interest is $\boldsymbol{\theta} = \left(\boldsymbol{\beta}^T, {\phi}\right)^T$ 
which is common across the densities.
Following \cite{Ghosh/Basu:2013}, we define the {MDPDE} of $\boldsymbol{\theta }$ 
under the BRM (\ref{EQ:BRM}) as the minimizer of the average DPD measure with tuning parameter $\alpha \geq 0$ given by 
\begin{eqnarray}
{n}^{-1}\sum^{n}_{i=1}d_{\alpha}\left(\widehat{g}_i\left(\cdot\right),f_i\left(\cdot, \boldsymbol{\theta }\right)\right),
\label{EQ:Avg_DPD}
\end{eqnarray}
where $\widehat{g}_i$ is an estimate of $g_i$ based on the given data.
Since the DPD measure is a proper statistical divergence, 
the resulting minimizer is clearly Fisher consistent for $\boldsymbol\theta$.
For the present case of BRM (\ref{EQ:BRM}), 
since we have only one observation from each $g_i$,
a simple estimate of it is given by the degenerate distribution at $y_i$ for any $i=1,\dots,n$.
Hence, after some simplification, the minimizer of (\ref{EQ:Avg_DPD}) 
is seen to be the minimizer of the simpler objective function {(Eq.(1) of the online supplement)}
\begin{eqnarray}
H_{n,\alpha}(\boldsymbol\theta) = n^{-1}\sum^n_{i=1}
\left[K_{i,\alpha}(\boldsymbol\theta) - \frac{1+\alpha}{\alpha}f_i(y_i, \boldsymbol\theta)^\alpha \right],
\label{EQ:Hn}
\end{eqnarray}
with  
$K_{i,\alpha }(\boldsymbol\theta)
=\frac{B\left((1+\alpha)\mu_i\phi -\alpha , (1+\alpha)\left(1-\mu_i\right)\phi -\alpha \right)}{
B{\left({\mu }_i\phi , \left(1- \mu_i\right)\phi \right)}^{\alpha}}.$
We need to minimize this objective function $H_{n,\alpha}(\boldsymbol\theta) $ with respect to 
$\boldsymbol\theta=(\boldsymbol\beta^T, \phi)^T$ 
to obtain its MDPDE with tuning parameter $\alpha$, 
say $\widehat{\boldsymbol\theta}_{n,\alpha}=(\widehat{\boldsymbol\beta}_{n,\alpha}^T, \widehat{\phi}_{n,\alpha})^T$.
Note that,  the above objective function $H_{n,\alpha}(\boldsymbol\theta)$ becomes [1 $-$ log-likelihood] as $\alpha \to 0$
and hence the proposed MDPDE at $\alpha=0$ coincides with the usual MLE of \cite{Ferrari/Cribari-Neto:2004}
which is known to be non-robust but fully efficient.
Further, {the} MDPDE at $\alpha=1$ coincides with the minimum $L_2$-distance estimator
which is known to be highly robust but inefficient under any general model. 
Hence the tuning parameter $\alpha$ in the proposed MDPDE under the BRM
is expected to yield a trade-off between robustness and efficiency of the estimator.

Equivalently, we can also obtain the MDPDE  
$\widehat{\boldsymbol\theta}_{n,\alpha}=(\widehat{\boldsymbol\beta}_{n,\alpha}^T, \widehat{\phi}_{n,\alpha})^T$
by solving the estimating equations obtained by differentiating the  objective function 
$H_{n,\alpha}(\boldsymbol\theta)$ with respect to $\boldsymbol{\theta}=(\boldsymbol{\beta}^T, \phi)^T$. 
For the BRM, these estimating equations simplify to
{(from Eq.(2) of the online supplement)}
\begin{eqnarray}
\sum^{n}_{i=1}\left[\gamma^{(\alpha)}_{1,i}(\boldsymbol\theta) - \left(y^*_{1,i} - \mu^*_{1,i}\right)
\frac{\phi}{g'(\mu_i)}f_i(y_i, \boldsymbol\theta)^\alpha\right]\boldsymbol{x}_i &=& {\boldsymbol{0}_p},\label{EQ:Est_Eqn1}\\
\sum^n_{i=1}\left[\gamma^{(\alpha)}_{2,i}(\boldsymbol\theta) 
- \left\{\mu_i\left(y^*_{1,i} - \mu^*_{1,i}\right)+ \left(y^*_{2,i} - \mu^*_{2,i}\right)
\right\}f_i(y_i, \boldsymbol\theta)^\alpha\right] &=& 0, \label{EQ:Est_Eqn2}
\end{eqnarray}
where {$\boldsymbol{0}_{p}$ is the zero vector of length $p$,} 
$g'$ denotes the derivative of $g$ and
\begin{eqnarray}
y^*_{1,i}&=&\log\frac{y_i}{1-y_i},
~~~~~
{\mu }^*_{1,i}=E\left(y^*_{1,i}\right)=\psi \left({\mu }_i\phi \right) - \psi \left(\left(1-{\mu }_i\right)\phi \right),
\nonumber\\
y^*_{2,i}&=&\mathrm{log} \left(1-y_i\right),
~~~~~~~~
{\mu }^*_{2,i}=E\left(y^*_{2,i}\right)=\psi \left(\left(1-{\mu }_i\right)\phi \right) - \psi \left(\phi \right),\nonumber\\
\gamma^{\left(\alpha \right)}_{1,i}(\boldsymbol\theta) 
&=& \left(\psi\left(a_{i,\alpha}\right) - \psi\left(b_{i,\alpha}\right) - \mu^*_{1,i}\right)
\frac{\phi K_{i,\alpha}(\boldsymbol\theta )}{g'(\mu_i)},
\nonumber\\
{\gamma }^{\left(\alpha \right)}_{2,i}\left(\boldsymbol\theta \right) 
&=& \left[ {\mu }_i\left(\psi\left(a_{i,\alpha }\right)-\psi \left(b_{i,\alpha }\right) - \mu^*_{1,i}\right)
+ \left(\psi\left(b_{i,\alpha }\right) -\psi\left(a_{i,\alpha } + b_{i,\alpha }\right) - \mu^*_{2,i}\right)\right]
K_{i,\alpha }(\boldsymbol\theta)
\nonumber
\end{eqnarray}
with $a_{i,\alpha }=\left(1+\alpha \right)\mu_i\phi -\alpha$, $b_{i,\alpha}=(1+\alpha)(1-\mu_i)\phi -\alpha$
and $\psi(\cdot)$ being the digamma function.
Clearly the estimating equations are unbiased at the model for any $\alpha\geq 0$.
Also, at $\alpha=0$, we have $\gamma^{\left(\alpha \right)}_{1,i}(\boldsymbol\theta) =0
=\gamma^{\left(\alpha \right)}_{2,i}(\boldsymbol\theta)$
for all $i=1, \ldots, n$ and these MDPDE estimating equations then coincide 
with the MLE estimating (score) equations as expected.

The asymptotic distribution of this proposed MDPDE can be derived from 
{the} general results of \cite{Ghosh/Basu:2013} under Assumptions (A1)--(A7) of their paper,
{also presented in the online supplement}. 
In particular, whenever the model assumption (\ref{EQ:BRM}) holds with true parameter value $\boldsymbol\theta_0$,
i.e., $g_i(\cdot) = f_i(\cdot, \boldsymbol\theta_0)$ for all $i$, we have the following
{from Result R1 of the online supplement.}
\begin{enumerate}
\item There exists a consistent sequence $\widehat{\boldsymbol\theta}_{n,\alpha }$ 
of roots to the estimating equations (\ref{EQ:Est_Eqn1}) and (\ref{EQ:Est_Eqn2}).
		
\item Asymptotically 
$\boldsymbol\Omega_n(\boldsymbol\theta_0)^{-1/2}\boldsymbol\Psi_n(\boldsymbol\theta_0) 
\left[\sqrt{n}\left(\widehat{\boldsymbol\theta}_{n, \alpha} - \boldsymbol\theta_0\right)\right]
\sim N_{p+1}\left(\boldsymbol{0}_{p+1}, \boldsymbol{I}_{p+1}\right)$,
where $\boldsymbol{I}_{p+1}$ is identity matrix of order $(p+1)$ and
\end{enumerate}
\begin{eqnarray}
\boldsymbol\Psi_n\left(\boldsymbol\theta \right)&=&\frac{1}{n}\sum^n_{i=1}\left[ \begin{array}{cc}
	{\gamma }^{\left(\alpha \right)}_{11,i}\left(\boldsymbol\theta \right)\boldsymbol{x}_i\boldsymbol{x}^T_i 
	& {\gamma }^{\left(\alpha \right)}_{12,i}\left(\boldsymbol\theta \right)\boldsymbol{x}_i \\ 
	{\gamma }^{\left(\alpha \right)}_{12,i}\left(\boldsymbol\theta \right)\boldsymbol{x}^T_i 
	& {\gamma }^{\left(\alpha \right)}_{22,i}\left(\boldsymbol\theta \right) \end{array}
	\right],\nonumber\\
\boldsymbol\Omega_n\left(\boldsymbol\theta \right)&=&\frac{1}{n}\sum^n_{i=1}\left[ \begin{array}{cc}
	\left\{{\gamma }^{\left(2\alpha \right)}_{11,i}\left(\boldsymbol\theta \right)-{\gamma }^{\left(\alpha \right)}_{1,i_0}{\left(\boldsymbol\theta \right)}^2\right\}\boldsymbol{x}_i\boldsymbol{x}^T_i 
	& \left\{{\gamma }^{\left(2\alpha \right)}_{12,i}\left(\boldsymbol\theta \right)-{\gamma }^{\left(\alpha \right)}_{1,i_0}\left(\boldsymbol\theta \right){\gamma }^{\left(\alpha \right)}_{2,i_0}\left(\boldsymbol\theta \right)\right\}\boldsymbol{x}_i \\ 
	\left\{{\gamma }^{\left(2\alpha \right)}_{12,i}\left(\boldsymbol\theta \right)-{\gamma }^{\left(\alpha \right)}_{1,i_0}\left(\boldsymbol\theta \right){\gamma }^{\left(\alpha \right)}_{2,i_0}\left(\boldsymbol\theta \right)\right\}\boldsymbol{x}^T_i 
	& \left\{{\gamma }^{\left(2\alpha \right)}_{22,i}\left(\boldsymbol\theta \right)-{\gamma }^{\left(\alpha \right)}_{2,i_0}{\left(\boldsymbol\theta \right)}^2\right\} \end{array}
	\right],\nonumber
\end{eqnarray}
with explicit forms of ${\gamma }^{\left(\alpha \right)}_{jk,i}\left(\boldsymbol\theta \right)$ being given by
\begin{eqnarray}
\gamma_{11, i}^{(\alpha)}(\boldsymbol\theta) &=& \frac{\phi^2K_{i,\alpha}(\boldsymbol\theta)}{g'(\mu_i)^2}
\left[\psi_1(a_{i,\alpha}) + \psi_1(b_{i,\alpha}) +(\psi(a_{i,\alpha}) - \psi(b_{i,\alpha}) - \mu_{1,i}^\ast)^2\right]
\nonumber\\
\gamma_{12, i}^{(\alpha)}(\boldsymbol\theta) &=& \frac{\phi K_{i,\alpha}(\boldsymbol\theta)}{g'(\mu_i)}
\left[\mu_i\left\{\psi_1(a_{i,\alpha}) + \psi_1(b_{i,\alpha}) 
+(\psi(a_{i,\alpha}) - \psi(b_{i,\alpha}) - \mu_{1,i}^\ast)^2\right\} \right.\nonumber\\
&&~~~~~~~~~~~~~\left. + \left\{ -\psi_1(b_{i,\alpha})+(\psi(a_{i,\alpha}) - \psi(b_{i,\alpha}) - \mu_{1,i}^\ast)
(\psi(b_{i,\alpha}) - \psi(a_{i,\alpha}+b_{i,\alpha}) - \mu_{2,i}^\ast)\right\}\right]
\nonumber\\
\gamma_{22, i}^{(\alpha)}(\boldsymbol\theta) &=& K_{i,\alpha}(\boldsymbol{\theta})
\left[\mu_i^2\left\{\psi_1(a_{i,\alpha}) + \psi_1(b_{i,\alpha}) 
+(\psi(a_{i,\alpha}) - \psi(b_{i,\alpha}) - \mu_{1,i}^\ast)^2\right\} \right.\nonumber\\
&&~~~~~~~~~~\left. + 2\mu_i\left\{ -\psi_1(b_{i,\alpha})+(\psi(a_{i,\alpha}) - \psi(b_{i,\alpha}) - \mu_{1,i}^\ast)
(\psi(b_{i,\alpha}) - \psi(a_{i,\alpha}+b_{i,\alpha}) - \mu_{2,i}^\ast)\right\}\right.\nonumber\\
&&~~~~~~~~~~\left. + \left\{ \psi_1(b_{i,\alpha}) - \psi_1(a_{i,\alpha}+b_{i,\alpha})
+(\psi(b_{i,\alpha}) - \psi(a_{i,\alpha}+b_{i,\alpha}) - \mu_{2,i}^\ast)^2\right\}
\right],\nonumber
\end{eqnarray}
and $\psi_1$ being the trigamma function.
The required conditions (A1)--(A7) of \cite{Ghosh/Basu:2013} can be verified to hold under mild boundedness conditions 
on the given covariate values (fixed design).
However, the form of the above asymptotic variance matrix indicates that, 
given any fixed design, the asymptotic relative efficiency of the proposed MDPDE decreases as $\alpha$ increases
but this loss in efficiency is not significant at small positive values of $\alpha$.
We will verify this property empirically again in Section {\ref{SEC:SimMDPDE}};
but this small loss in efficiency leads to increased robustness of the proposed estimator
over the non-robust MLE which we justify through the influence function analysis in the next subsection.

\subsection{Influence Function of the MDPDE under the BRM}
\label{SEC:MDPDE_IF}
 
The influence function {(IF)} is a classical tool to measure the theoretical robustness property 
of any estimator under the iid set-up \citep{Hampel/etc:1986}. 
It measures the asymptotic bias due to infinitesimal contamination in the data.
The concept has been suitably extended and applied to the {case} of non-homogeneous observations by 
\cite{Huber:1983,Ghosh/Basu:2013,Ghosh/Basu:2016,Aerts/Haesbroeck:2016},
where the corresponding statistical functional and the {IF} both depend on the sample size $n$ 
(unlike the iid case).
Note that, for such non-homogeneous cases the contamination can be in any of the distributions indexed by $i$ or in all of them.
We use this concept to illustrate the robustness of our proposed  MDPDE under the BRM.

Assuming $G_i$ to be the true distribution function of $y_i$ corresponding to the density $g_i$ for each $i$,
the statistical functional corresponding to the MDPDE of $\boldsymbol\theta$ under the BRM  (\ref{EQ:BRM}) is defined as
\begin{eqnarray}
\boldsymbol{T}_\alpha(G_1, \ldots, G_n) = \arg\min_{\boldsymbol\theta} {n}^{-1}\sum^{n}_{i=1}
d_{\alpha}\left(g_i(\cdot),f_i\left(\cdot,\boldsymbol{\theta }\right)\right),
\label{EQ:DPD_func}
\end{eqnarray}
whenever the minimum exists. This is a Fisher consistent functional at the assumed BRM by the definition of the DPD measure.
Suppose first, for simplicity,  the contamination is in only the $i_0$-th distribution
through	$G_{i_0,\epsilon }=\left(1-\epsilon \right)G_{i_0}+\epsilon {\wedge }_{t_{i_0}}$, 
where $\epsilon $ is the contamination proportion and 
${\wedge }_{t_{i_0}}$ is the degenerate distribution at the contamination point $t_{i_0}$.
The corresponding (first order) influence function (IF) of the proposed MDPDE functional $\boldsymbol{T}_\alpha$ 
is defined as
\begin{eqnarray}
\mathcal{IF}\left(t_{i_0},\boldsymbol{T}_\alpha; G_1, \dots, G_n\right) &=& 
\left|\frac{\partial \boldsymbol{T}_\alpha(G_1, \dots, G_{i_0,\epsilon}, \dots, G_n)}{\partial\epsilon}\right|_{\epsilon =0} 
\nonumber\\
&=& \lim\limits_{\epsilon\downarrow 0} 
\frac{\boldsymbol{T}_\alpha(G_1, \dots, G_{i_0,\epsilon}, \dots, G_n) - \boldsymbol{T}_\alpha(G_1, \dots, G_n)}{\epsilon}. 
\nonumber
\end{eqnarray}	
Note that, whenever this IF is bounded in $t_{i_0}$, 
the asymptotic bias due to infinitesimal contamination at $G_{i_0}$ remains bounded, 
implying the robustness of the corresponding estimator. 
On the other hand, if this IF is unbounded in $t_{i_0}$, 
then the same bias may tend to infinity for distant contaminations
implying the non-robust	nature of the estimator. 

For our beta regression model with $g_i(\cdot) = f_i\left(\cdot, \boldsymbol\theta \right)$ for all $i$,
some calculations{, based on the general Result R2(i) of the online supplement,} 
yield the simplified form of the above IF as given by 
\begin{eqnarray}
\mathcal{IF}\left(t_{i_0},\boldsymbol{T}_\alpha; F_1, \dots, F_n\right) 
= \boldsymbol\Psi_n\left(\boldsymbol\theta\right)^{-1}
\left[ \begin{array}{c}
\left({{t}}^{{*}}_{{1},{{i}}_{{0}}}~{-}{{\mu }}^{{*}}_{{1},{{i}}_{{0}}}\right){\frac{{\phi }}{g'\left({{\mu }}_{{{i}}_{{0}}}\right)}}f_{i_0}{\left({{t}}_{{{i}}_{{0}}}, \boldsymbol{\theta }\right)}^{{\alpha }}
-{{\gamma }}^{\left({\alpha }\right)}_{{1},{{i}}_{{0}}}\left(\boldsymbol{\theta }\right) \\ 
\left\{{{\mu }}_{{i}}\left({{t}}^{{*}}_{{1},{{i}}_{{0}}}~{-}{{\mu }}^{{*}}_{{1},{{i}}_{{0}}}\right){+}\left({{t}}^{{*}}_{{2},{{i}}_{{0}}}~{-}{{\mu }}^{{*}}_{{2},{{i}}_{{0}}}\right)\right\}{{f}}_{{{i}}_{{0}}}{\left({{t}}_{{{i}}_{{0}}}, \boldsymbol{\theta }\right)}^{{\alpha }}
-{\gamma }^{\left({\alpha }\right)}_{{2},{{i}}_{{0}}}\left(\boldsymbol{\theta }\right) \end{array}
\right],\nonumber
\end{eqnarray}
where $t^*_{1,i_0} = \log\frac{t_{i_0}}{1-t_{i_0}}$, $t^*_{2,i}=\log\left(1-t_{i_0}\right)$
and $F_i$ is the distribution function of $f_i(\cdot, \boldsymbol\theta)$ for each $i=1, \dots, n$.
Clearly this IF of the proposed MDPDE is bounded for all $\alpha>0$ but unbounded at $\alpha = 0$.
This implies that the proposed MDPDE with $\alpha>0$ is robust against contamination in data, 
whereas that at $\alpha=0$ (existing MLE) is clearly non-robust.
Further, it can also be verified that, given any fixed design, 
the supremum of this IF decreases as $\alpha$ increases, 
which in turn implies the increase in their robustness.
This fact will be further seconded through empirical illustrations in Section {\ref{SEC:SimMDPDE}}.
 
Similar results can also be obtained if there are contaminations in all the $G_i$s
{(see Result R2(ii) of the online supplement)}.
The resulting influence function is then the sum of the previous IFs for individual component-wise contaminations
and hence the implication is again the same indicating robustness at $\alpha>0$ and non-robustness at $\alpha=0$.

\section{Robust Hypothesis Testing: A Wald-Type Test Statistics}
\label{SEC:test}

Let us now consider the second important aspect of statistical inference, namely the testing of statistical hypothesis.
As noted previously, the existing MLE based likelihood ratio tests or Wald tests 
are highly non-robust against data contamination in any \trd{general non-homogeneous} set-up including the BRM. 
Suitable robust hypothesis testing procedures under the general non-homogeneous set-up 
have been developed in \cite{Ghosh/Basu:2017} and \cite{Basu/etc:2017} by extending 
the likelihood ratio and the Wald-type tests respectively.
In this section, we develop a robust hypothesis testing procedure based on the proposed MDPDE for the BRM; 
here we restrict ourselves only to the Wald-type tests which are easy to implement in practice. 
{Related background results from \cite{Basu/etc:2017} are again provided in the online supplement for the sake of completeness.}

Consider the BRM (\ref{EQ:BRM}) with the set-up as discussed in the previous sections.
Consider the most common class of general linear hypotheses given by
\begin{eqnarray}
H_0 : \boldsymbol{M}\boldsymbol\beta= \boldsymbol{m}_0~~~\mbox{against }~~~~
H_1 : \boldsymbol{M}\boldsymbol\beta \neq \boldsymbol{m}_0,
\label{EQ:GLHyp}
\end{eqnarray}
where $\boldsymbol{M}$ is a known matrix of order $r\times p$ and $\boldsymbol{m}_0$ is a known $r$-vector of reals. 
We make the standard assumption that $rank\left(\boldsymbol{M}\right)=r$ 
so that there exists a true null parameter value $\boldsymbol\beta_0\neq \boldsymbol{0}_p$ (say)
satisfying $\boldsymbol{M}\boldsymbol\beta_0=\boldsymbol{m}_0$. Suppose 
$\widehat{\boldsymbol\theta}_{n,\alpha }=\left(\widehat{\boldsymbol\beta }_{n,\alpha}^T, \widehat{\phi}_{n,\alpha}\right)^T$
denotes the MDPDE of $\boldsymbol\theta=(\boldsymbol\beta^T, \phi)^T$ under the BRM (\ref{EQ:BRM}).
We define the Wald-Type test statistic for testing hypothesis (\ref{EQ:GLHyp})
as
\begin{eqnarray}
W_{n,\alpha} =n\left(\boldsymbol{M}\widehat{\boldsymbol\beta}_{n,\alpha} - \boldsymbol{m}_0\right)^T\left[\boldsymbol{M}
\boldsymbol\Psi_n^{11}(\widehat{\boldsymbol\theta}_{n, \alpha})^{-1}
\boldsymbol\Omega_n^{11}(\widehat{\boldsymbol\theta}_{n,\alpha})
\boldsymbol\Psi_n^{11}(\widehat{\boldsymbol\theta}_{n, \alpha})^{-1}
\boldsymbol{M}^T\right]^{-1}\left(\boldsymbol{M}\widehat{\boldsymbol\beta}_{n,\alpha} - \boldsymbol{m}_0\right), 
\label{EQ:WaldTS}
\end{eqnarray}
where $\boldsymbol\Psi_n^{11}$ and $\boldsymbol\Omega_n^{11}$ 
are the $p\times p$ {principal} sub-matrix of the matrices $\boldsymbol\Psi_n$ and $\boldsymbol\Omega_n$ respectively
and are given by 
$\boldsymbol\Psi_n^{11}\left(\boldsymbol\theta \right)
= \frac{1}{n}\sum^n_{i=1}{\gamma }^{\left(\alpha \right)}_{11,i}\left(\boldsymbol\theta \right)
\boldsymbol{x}_i\boldsymbol{x}^T_i$
and 
$\boldsymbol\Omega_n^{11}\left(\boldsymbol\theta\right)
=\frac{1}{n}\sum^n_{i=1}\left\{{\gamma }^{\left(2\alpha \right)}_{11,i}\left(\boldsymbol\theta \right)
-{\gamma }^{\left(\alpha \right)}_{1,i_0}{\left(\boldsymbol\theta \right)}^2\right\}\boldsymbol{x}_i\boldsymbol{x}^T_i$.
Note that, since the MDPDE at $\alpha=0$ coincides with the MLE, 
the test statistic $W_{n, 0}$ is {nothing} but the non-robust MLE based {classical} Wald test.
So, the proposed test statistics $W_{n,\alpha}$ are the robust generalization of the Wald tests
and hence referred to as the Wald-type tests.

In particular, for testing the significance of individual regression coefficient $\beta_j$, i.e., testing  
\begin{eqnarray}
H_0 : \beta_j= 0~~~\mbox{against }~~~~H_1 : \beta_j \neq 0,
\label{EQ:IndHyp}
\end{eqnarray}
for any $j=1,\dots, p$, the proposed test statistic (\ref{EQ:WaldTS}) simplifies to 
$W_{n,\alpha} = \frac{n\widehat{\beta}_{n,\alpha,j}^2}{{\sigma}_j(\widehat{\boldsymbol\beta}_{n,\alpha})}$,
where $\widehat{\beta}_{n,\alpha,j}$ is the MDPDE of $\beta_j$ and ${\sigma}_j(\boldsymbol\theta)$ 
is the asymptotic variance of $\sqrt{n}\widehat{\beta}_{n,\alpha,j}$.

\subsection{Asymptotic Properties}

The first property that we need for any proposed test statistic is its null distribution 
to find out the critical region of the test.
Although {the} exact null distribution is not easy to obtain in general, 
the asymptotic distribution of our proposed test statistic $W_{n, \alpha} $ 
can be derived directly from that of the MDPDE.
We assume that the matrices involved in the asymptotic variance of the MDPDE of $\boldsymbol\beta$, 
namely $\boldsymbol\Psi_n^{11}$ and $\boldsymbol\Omega_n^{11}${,} are continuous in $\boldsymbol\theta$.
Then, it is straightforward from the results of Section \ref{SEC:MDPDE_beta}
that the asymptotic null distribution of $W_{n,\alpha}$ for hypothesis (\ref{EQ:GLHyp}) 
is $\chi^2_r$, the chi-square distribution with $r$ degrees of freedom
{(see Result R3(i) in the online supplement)}.
So, the critical region of the proposed testing procedure at $\alpha_0$-level of significance is given by 
{$\left\{W_{n,\alpha} > \chi^2_{r,\alpha_0} \right\},$}
where $\chi^2_{r,\alpha_0}$ is the $(1-\alpha_0)$-th quantile of the $\chi_r^2$ distribution.
For the particular case of the hypothesis (\ref{EQ:IndHyp}), 
the corresponding null asymptotic distribution of $W_{n,\alpha}$ is $\chi_1^2$. 
So, we can also perform the one-sided testing for the significance of $\beta_j$ by considering the {test statistic}
$W_{n,\alpha}^{+} = \frac{\sqrt{n}\widehat{\beta}_{n,\alpha,j}}{\sqrt{{\sigma}_j(\widehat{\boldsymbol\beta}_{n,\alpha})}}$,
which has an asymptotic standard normal distribution at the null hypothesis in (\ref{EQ:IndHyp}).

Further we can apply suitable results from \cite{Basu/etc:2017} on the Wald-type tests for the general non-homogeneous set-up
to obtain useful power approximations for our proposal in the BRM. In particular, 
{by Result R3(ii) of the online supplement,}
the tests based on $W_{n,\alpha}$ are consistent at any fixed alternative for every $\alpha\geq 0$;
this fact also follows from the Fisher consistency of the {MDPDEs} used in the construction of test statistics 
and we leave the details for the reader.

So, for the {purpose of comparison}, we need to compute the asymptotic power under the contiguous sequence of alternatives
$H_{1,n} : \boldsymbol\beta_n = \boldsymbol\beta_0 + \frac{\boldsymbol{d}}{\sqrt{n}}$ 
for $\boldsymbol{d} \in \mathbb{R}^p -\{\boldsymbol{0}_p\}$,
where $\boldsymbol\theta_0=(\boldsymbol\beta_0^T, \phi_0)^T$ is the null parameter value satisfying 
$\boldsymbol{M}\boldsymbol\beta_0=\boldsymbol{m}_0$.
However, using the asymptotic distribution of the MDPDE from Section \ref{SEC:MDPDE_beta},
one can obtain the asymptotic distribution of our test statistics $W_{n,\alpha}$ under the hypothesis $H_{1,n}$ 
to be $\chi^2_r(\delta)$, the non-central ${\chi }^2$ with degrees of freedom $r$ and non-centrality parameter 
$
\delta =\boldsymbol{d}^T\boldsymbol{M} \left[\boldsymbol{M}
\boldsymbol\Psi_n^{11}(\boldsymbol\theta_0)^{-1}\boldsymbol\Omega_n^{11}(\boldsymbol\theta_0)
\boldsymbol\Psi_n^{11}(\boldsymbol\theta_0)^{-1}\boldsymbol{M}^T\right]^{-1}\boldsymbol{M}^T\boldsymbol{d}
$
{(see Result R3(iii) of the online supplement)}.
The asymptotic contiguous power of the proposed testing procedure can then be computed as 
$\left[1-F_{\chi^2_r(\delta)}(\chi^2_{r,\alpha_0})\right]$, 
where $F_{\chi^2_r(\delta)}$ denotes the distribution function of $\chi^2_r(\delta)$.
In particular, the pitman's asymptotic relative efficiencies of \trd{the} $W_{n,\alpha}$ based Wald-type tests at $\alpha>0$
with respect to the most powerful (but non-robust) classical Wald test (at $\alpha=0$)
depend on the non-centrality parameter $\delta$ and {are} directly proportional to the ratio of the inverse variance 
matrix of the  MDPDE {and} the MLE. Hence, {they are indeed} directly proportional to the asymptotic efficiency of the MDPDE itself. 
So, for any given fixed design, the asymptotic power under contiguous alternative decreases slightly with increasing $\alpha$
but the loss is not quite significant at small positive $\alpha$ as in the case of efficiency of the MDPDEs;
see Section {\ref{SEC:SimWald}} for corresponding empirical illustrations.

\subsection{Robustness Analysis}

We theoretically study the robustness of the proposed Wald-type tests
through the corresponding influence function analysis \citep{Hampel/etc:1986}. 
Considering the set-up of Section \ref{SEC:MDPDE_IF}, 
we define the statistical functional corresponding to the  proposed test statistics $W_{n,\alpha}$
(ignoring the multiplier $n$) as 
\begin{eqnarray}
W_\alpha(G_1, \ldots, G_n) = \left(\boldsymbol{M}\boldsymbol{T}_\alpha(G_1, \ldots, G_n) - \boldsymbol{m}_0\right)^T
\boldsymbol\Sigma(G_1, \ldots, G_n)^{-1}\left(\boldsymbol{M}\boldsymbol{T}_\alpha(G_1, \ldots, G_n) - \boldsymbol{m}_0\right), 
\end{eqnarray}
where 
$ \boldsymbol\Sigma(G_1, \ldots, G_n) = \left[\boldsymbol{M}
\boldsymbol\Psi_n^{11}(\boldsymbol{T}_\alpha(G_1, \ldots, G_n))^{-1}
\boldsymbol\Omega_n^{11}(\boldsymbol{T}_\alpha(G_1, \ldots, G_n))
\boldsymbol\Psi_n^{11}(\boldsymbol{T}_\alpha(G_1, \ldots, G_n))^{-1}\boldsymbol{M}^T\right]$ 
and $\boldsymbol{T}_\alpha(G_1, \ldots, G_n)$ is the functional for the MDPDE as defined in (\ref{EQ:DPD_func}).
We can define its influence function as in the case of estimation
by assuming contamination in any fixed distribution or in all distributions.

Let us again consider the contamination only in the distribution $G_{i_0}$ at the contamination point $t_{i_0}$.
Then, using the Fisher consistency of $\boldsymbol{T}_\alpha$, a routine differentiation yields 
the (first order) influence function of the test functional $W_\alpha$ to be identically zero at the model,
i.e., 
\begin{eqnarray}
\mathcal{IF}\left(t_{i_0}, W_\alpha; F_1, \dots, F_n\right) = 0.\nonumber
\end{eqnarray}
Therefore, this first order influence function cannot indicate the robustness of our proposed Wald-type tests,
which is expected from the literature of similar quadratic tests 
\citep{Heritier/Ronchetti:1994, Toma/Broniatowski:2010, Ghosh/Basu:2017, Basu/etc:2017}.
So, we need to consider the second order influence function for $W_\alpha$
defined analogously with the second order partial derivative as
\begin{eqnarray}
\mathcal{IF}_2\left(t_{i_0},W_\alpha; G_1, \dots, G_n\right) &=& 
\left|\frac{\partial^2 W_\alpha(G_1, \dots, G_{i_0,\epsilon}, \dots, G_n)}{\partial^2\epsilon}\right|_{\epsilon =0}.
\nonumber
\end{eqnarray}	
It indicates a second order approximation to the asymptotic bias due to infinitesimal contamination
in contrast to the first order approximation provided by the first order IF. 
For the present BRM some calculations{, based on Result R4(i) of the online supplement,} 
yield the form of this second order IF for the proposed Wald-type test
functional $W_\alpha$ at the model as given by
\begin{eqnarray}
\mathcal{IF}_2\left(t_{i_0}, W_\alpha; F_1, \dots, F_n\right) = 
\mathcal{IF}\left(t_{i_0}, \boldsymbol{T}_\alpha; F_1, \dots, F_n\right)\boldsymbol{M}^T
\boldsymbol\Sigma(F_1, \ldots, F_n)^{-1}\boldsymbol{M}\mathcal{IF}\left(t_{i_0}, \boldsymbol{T}_\alpha; F_1, \dots, F_n\right).
\nonumber
\end{eqnarray}
Therefore, this influence function is bounded if and only if the IF of the MDPDE $\boldsymbol{T}_\alpha$,
derived in Section \ref{SEC:MDPDE_IF}, is bounded and this holds only for all $\alpha>0$. 
Hence, the proposed Wald-type test statistics are expected to be robust  for $\alpha>0$
but non-robust at $\alpha=0$ (which is the classical MLE based Wald test);
further numerical illustrations are given in Section {\ref{SEC:SimWald}}.

We can also examine the influence of the contamination on the level and power 
of the proposed Wald-type tests through the level and power influence function analysis 
\citep{Hampel/etc:1986, Ghosh/Basu:2017}. For this purpose, we can directly apply 
the corresponding results for the general non-homogeneous cases from \cite{Basu/etc:2017},
{described in Result R4(ii) of the online supplement},
to conclude that the power influence function is indeed a matrix multiple of the 
IF of the MDPDE $\boldsymbol{T}_\alpha$. Therefore, the proposed test is robust in asymptotic contiguous power
whenever the IF of $\boldsymbol{T}_\alpha$ is bounded, i.e., for all $\alpha>0$, but is non-robust at $\alpha=0$.
However, following {the same result }\cite{Basu/etc:2017}, the level influence function of this type of tests 
under non-homogeneous data is {identically} zero {whenever the IF of $\boldsymbol{T}_\alpha$ is bounded,}
indicating the robustness of asymptotic level for all {$\alpha> 0$}
against infinitesimal contiguous contamination at the null hypothesis.

	\section{\textcolor{black}{Simulation Studies}}
\label{SEC:Simulation}
\subsection{\textcolor{black}{Performance of the MDPDE}}\label{SEC:SimMDPDE}
Let us now study the finite-sample behavior of the proposed estimator, MDPDE, 
through suitable simulation {studies}  and  compare {them} with theoretical (asymptotic) results.	
Consider {a} sample size {$n$} and fix $n$ covariate values $x_1, \dots, x_n$ being independent observations from $U(0, 1)$.	
We generate 1000 samples from the BRM (\ref{EQ:BRM}) with $p=2$, one intercept ($\beta_1$) 
and one slope ($\beta_2$) corresponding to the covariates $x_i$,
along with the logit link function. The true value of the parameter $\boldsymbol{\theta}=(\beta_1, \beta_2, \phi)^T$
is taken as $(-1, 1, 5)^T$. For each of the samples, we compute the MDPDEs with different $\alpha$
and derive their empirical bias and MSE over these 1000 replications (without any outlier){; the results}
are reported in Table \ref{TAB:pure_data} {for sample sizes $n=50, 100$}.
Clearly, MLE has the minimum absolute bias and MSE under pure data as expected
and the bias and MSE of the proposed MDPDE increase slightly with increasing $\alpha$.
But this increase in bias or MSE is not quite significant at small positive $\alpha $ like 0.3, 0.4,
which is consistent with the asymptotic efficiency described in Section \ref{SEC:MDPDE_beta}.

\begin{table}[h]
	\caption{Empirical Bias and MSE of the MDPDEs with different $\alpha$ under pure data}
	\centering	
{	
	\resizebox{\textwidth}{!}{
	\begin{tabular}{l|rrr|rrr|rrr|rrr}\hline
&	\multicolumn{6}{c|}{$n=50$}	&	\multicolumn{6}{c}{$n=100$}\\\hline
&	\multicolumn{3}{c|}{Bias}	&	\multicolumn{3}{c|}{MSE}	&	\multicolumn{3}{c|}{Bias}	&	\multicolumn{3}{c}{MSE}\\
		$\alpha$	&	$\beta_1$	&	$\beta_2$	& $\phi$	&	$\beta_1$	&	$\beta_2$	&	$\phi$ 
			&	$\beta_1$	&	$\beta_2$	& $\phi$	&	$\beta_1$	&	$\beta_2$	&	$\phi$\\\hline
		0 (MLE)		&	-0.010	&	0.011	&	0.332	&	0.058	&	0.172	&	1.124	&	-0.0042	&	0.004	&	0.202	&	0.032	&	0.097	&	0.537	\\
		0.1	&	-0.010	&	0.011	&	0.325	&	0.058	&	0.174	&	1.123	&	-0.0033	&	0.003	&	0.196	&	0.032	&	0.097	&	0.541	\\
		0.2	&	-0.012	&	0.013	&	0.338	&	0.059	&	0.178	&	1.180	&	-0.003	&	0.003	&	0.200	&	0.033	&	0.099	&	0.566	\\
		0.3	&	-0.014	&	0.015	&	0.367	&	0.062	&	0.185	&	1.293	&	-0.0031	&	0.003	&	0.211	&	0.034	&	0.102	&	0.607	\\
		0.4	&	-0.017	&	0.018	&	0.410	&	0.064	&	0.193	&	1.464	&	-0.0036	&	0.004	&	0.228	&	0.035	&	0.106	&	0.663	\\
		0.5	&	-0.020	&	0.022	&	0.464	&	0.067	&	0.203	&	1.696	&	-0.0042	&	0.004	&	0.248	&	0.037	&	0.110	&	0.731	\\
		0.6	&	-0.024	&	0.026	&	0.526	&	0.071	&	0.214	&	1.990	&	-0.005	&	0.005	&	0.271	&	0.038	&	0.115	&	0.810	\\
		0.7	&	-0.028	&	0.031	&	0.593	&	0.074	&	0.225	&	2.347	&	-0.0059	&	0.006	&	0.296	&	0.040	&	0.120	&	0.895	\\
		\hline
	\end{tabular}}}
	\label{TAB:pure_data}
\end{table}

\begin{table}[h]
	\caption{Empirical Bias and MSE of MDPDEs with different $\alpha$ under {contamination scheme (I)}}
	\centering	
	{	
		\resizebox{\textwidth}{!}{
			\begin{tabular}{l|rrr|rrr|rrr|rrr}\hline
				&	\multicolumn{6}{c|}{$n=50$}	&	\multicolumn{6}{c}{$n=100$}\\\hline
				&	\multicolumn{3}{c|}{Bias}	&	\multicolumn{3}{c|}{MSE}	&	\multicolumn{3}{c|}{Bias}	&	\multicolumn{3}{c}{MSE}\\
				$\alpha$	&	$\beta_1$	&	$\beta_2$	& $\phi$	&	$\beta_1$	&	$\beta_2$	&	$\phi$ 
				&	$\beta_1$	&	$\beta_2$	& $\phi$	&	$\beta_1$	&	$\beta_2$	&	$\phi$\\\hline
				0 (MLE)	&	0.232	&	-0.240	&	-0.332	&	0.141	&	0.283	&	1.066	&	0.2042	&	-0.192	&	-0.559	&	0.082	&	0.133	&	0.693	\\
				0.1	&	0.216	&	-0.218	&	-0.300	&	0.132	&	0.270	&	1.037	&	0.1845	&	-0.166	&	-0.508	&	0.074	&	0.122	&	0.644	\\
				0.2	&	0.199	&	-0.197	&	-0.247	&	0.125	&	0.262	&	1.050	&	0.1665	&	-0.144	&	-0.452	&	0.067	&	0.116	&	0.616	\\
				0.3	&	0.184	&	-0.177	&	-0.182	&	0.121	&	0.259	&	1.100	&	0.1511	&	-0.125	&	-0.397	&	0.063	&	0.113	&	0.607	\\
				0.4	&	0.170	&	-0.160	&	-0.111	&	0.119	&	0.260	&	1.196	&	0.1381	&	-0.109	&	-0.347	&	0.060	&	0.111	&	0.613	\\
				0.5	&	0.158	&	-0.145	&	-0.036	&	0.118	&	0.264	&	1.343	&	0.1272	&	-0.096	&	-0.301	&	0.059	&	0.112	&	0.630	\\
				0.6	&	0.147	&	-0.133	&	0.043	&	0.119	&	0.269	&	1.550	&	0.1181	&	-0.085	&	-0.259	&	0.058	&	0.113	&	0.656	\\
				0.7	&	0.138	&	-0.122	&	0.120	&	0.120	&	0.276	&	1.790	&	0.1104	&	-0.076	&	-0.220	&	0.057	&	0.115	&	0.689	\\
				\hline
	\end{tabular}}}
	\label{TAB:cont_data}
\end{table}

\begin{table}[h]
		
	\caption{Empirical Bias and MSE of MDPDEs with different $\alpha$ under contamination scheme (II)}
	\centering	
	\resizebox{\textwidth}{!}{
		\begin{tabular}{l|rrr|rrr|rrr|rrr}\hline
			&	\multicolumn{6}{c|}{$n=50$}	&	\multicolumn{6}{c}{$n=100$}\\\hline
			&	\multicolumn{3}{c|}{Bias}	&	\multicolumn{3}{c|}{MSE}	&	\multicolumn{3}{c|}{Bias}	&	\multicolumn{3}{c}{MSE}\\
			$\alpha$	&	$\beta_1$	&	$\beta_2$	& $\phi$	&	$\beta_1$	&	$\beta_2$	&	$\phi$ 
			&	$\beta_1$	&	$\beta_2$	& $\phi$	&	$\beta_1$	&	$\beta_2$	&	$\phi$\\\hline
			0 (MLE)	&	1.120	&	-1.588	&	-2.382	&	1.292	&	2.635	&	5.772	&	0.9779	&	-1.331	&	-2.271	&	0.977	&	1.829	&	5.205	\\
			0.1	&	1.067	&	-1.493	&	-2.348	&	1.186	&	2.363	&	5.612	&	0.8868	&	-1.193	&	-2.177	&	0.816	&	1.498	&	4.798	\\
			0.2	&	0.869	&	-1.203	&	-1.886	&	0.956	&	1.873	&	5.145	&	0.6326	&	-0.841	&	-1.681	&	0.509	&	0.925	&	3.392	\\
			0.3	&	0.450	&	-0.619	&	-0.690	&	0.575	&	1.136	&	3.635	&	0.2536	&	-0.335	&	-0.588	&	0.187	&	0.352	&	1.521	\\
			0.4	&	0.222	&	-0.307	&	-0.032	&	0.339	&	0.693	&	2.696	&	0.11	&	-0.146	&	-0.087	&	0.083	&	0.171	&	0.897	\\
			0.5	&	0.126	&	-0.175	&	0.243	&	0.216	&	0.464	&	2.301	&	0.0753	&	-0.101	&	0.065	&	0.059	&	0.131	&	0.778	\\
			0.6	&	0.092	&	-0.132	&	0.361	&	0.164	&	0.371	&	2.216	&	0.0686	&	-0.093	&	0.115	&	0.053	&	0.121	&	0.761	\\
			0.7	&	0.077	&	-0.114	&	0.426	&	0.130	&	0.309	&	2.250	&	0.0716	&	-0.097	&	0.129	&	0.052	&	0.121	&	0.782	\\
			\hline
	\end{tabular}}
	\label{TAB:cont_data2}
\end{table}

Next, to study the finite-sample robustness behavior of the proposed MDPDEs, 
we repeat the previous simulation study, but after contaminating each sample {through two different schemes}.
{In the contamination scheme (I),} we have randomly changed 10\% of the response values $y$ to $\left(1-y\right)$
and recalculated the MDPDEs based on the contaminated samples.
The corresponding bias and MSE are reported in Table \ref{TAB:cont_data}.
{Following the suggestion of a referee, in the second contamination scheme (II),
	we have replaced 5\% of the response ($y$) values associated with the minimum $x$-values to the extreme point $y=0.99$;
	the empirical bias and MSE for this extreme case of contamination are reported in Table \ref{TAB:cont_data2}.}
It can be observed that the absolute bias and MSE of the MLE are the worst,
since it is the most non-robust one. As $\alpha $ increases, 
both the absolute bias and MSE decrease significantly providing more accurate results{;
these become more prominent in the extreme contamination scheme (II).}	
Thus, the robustness of the proposed MDPDE under contamination significantly improves with increasing values of $\alpha$;
this is again consistent with the theoretical influence function analysis discussed in Section \ref{SEC:MDPDE_IF}.

Similar results are observed in several simulation studies with different contamination scheme and 
{different sample sizes}; so those are not repeated here for brevity. 
\\

{
	\subsection{\textcolor{black}{Performance of the Wald-type tests}}
		\label{SEC:SimWald}
}

In this section, we illustrate the empirical levels and powers of the proposed Wald-type tests based on the MDPDEs 
through simulation studies.  For the sake of consistency, 
let us consider the same simulation set-up as described in the previous section;
with each simulated sample of size {$n=50$ or 100}, both with and without contamination as before, 
we apply the proposed testing procedure for different hypotheses.
In particular, we perform the Wald-type tests with different $\alpha$ for six null hypotheses given by 
\begin{eqnarray}
& H_0^{L1} : \beta_1 = -1; ~~~~ H_0^{L2} : \beta_2 = 1; ~~~~ H_0^{L3} : (\beta_1, \beta_2)^T = (-1, 1)^T; ~~~~&
\mbox{for studying level,}\nonumber\\
\mbox{and }~~~~ & H_0^{P1} : \beta_1 = 0; ~~~~ H_0^{P2} : \beta_2 = 0; ~~~~ H_0^{P3} : (\beta_1, \beta_2)^T = (0, 0)^T; ~~~~&
\mbox{for studying power,}\nonumber
\end{eqnarray}
against their respective omnibus alternatives. Note that, all these hypotheses belong to the class
of general linear hypotheses (\ref{EQ:GLHyp}) considered in Section \ref{SEC:test}.
Based on 1000 replications, we compute the empirical {levels} and powers at the 5\% level of significance
for testing these hypotheses under pure data as well as under contaminated data; 
the results are reported in Tables {\ref{TAB:testpure_data}, \ref{TAB:testcont_data} and \ref{TAB:testcont_data2}} respectively.  

\begin{table}[h]
	\caption{Empirical {levels} and powers for the MDPDE based Wald-type tests for different null hypotheses 
		and different $\alpha$ under pure data}
	\centering	
{		
	\resizebox{\textwidth}{!}{
	\begin{tabular}{l|rrr|rrr|rrr|rrr}\hline
&	\multicolumn{6}{c|}{$n=50$}	&	\multicolumn{6}{c}{$n=100$}\\\hline
&	\multicolumn{3}{c|}{Size}	&	\multicolumn{3}{c|}{Power} &	\multicolumn{3}{c|}{Size}	&	\multicolumn{3}{c}{Power}\\
		$\alpha$	&	$H_0^{L1}$	&	$H_0^{L2}$	& $H_0^{L3}$	&	$H_0^{P1}$	&	$H_0^{P2}$	&	$H_0^{P3}$ 
			&	$H_0^{L1}$	&	$H_0^{L2}$	& $H_0^{L3}$	&	$H_0^{P1}$	&	$H_0^{P2}$	&	$H_0^{P3}$ \\\hline
		0 (Wald)	&	0.060	&	0.058	&	0.055	&	0.996	&	0.719	&	0.995	&	0.043	&	0.044	&	0.061	&	1.000	&	0.956	&	1.000	\\
		0.1	&	0.058	&	0.057	&	0.057	&	0.995	&	0.712	&	0.996	&	0.041	&	0.044	&	0.058	&	1.000	&	0.957	&	1.000	\\
		0.2	&	0.058	&	0.054	&	0.059	&	0.994	&	0.711	&	0.995	&	0.043	&	0.049	&	0.056	&	1.000	&	0.952	&	1.000	\\
		0.3	&	0.060	&	0.058	&	0.066	&	0.994	&	0.697	&	0.995	&	0.047	&	0.048	&	0.057	&	1.000	&	0.945	&	1.000	\\
		0.4	&	0.066	&	0.063	&	0.069	&	0.993	&	0.691	&	0.995	&	0.050	&	0.052	&	0.058	&	1.000	&	0.934	&	1.000	\\
		0.5	&	0.071	&	0.066	&	0.074	&	0.992	&	0.679	&	0.993	&	0.053	&	0.050	&	0.059	&	1.000	&	0.929	&	1.000	\\
		0.6	&	0.076	&	0.070	&	0.082	&	0.989	&	0.667	&	0.993	&	0.055	&	0.054	&	0.057	&	1.000	&	0.925	&	1.000	\\
		0.7	&	0.076	&	0.069	&	0.086	&	0.986	&	0.657	&	0.991	&	0.055	&	0.054	&	0.061	&	1.000	&	0.958	&	1.000	\\
		\hline
	\end{tabular}}}
	\label{TAB:testpure_data}
\end{table}

It can be observed that the {levels} of the MDPDE based Wald-type tests increase slightly under pure data for any hypothesis. 
In fact, {most of} the empirical {levels} are slightly inflated due to 
the use of asymptotic critical values for testing with {finite sample sizes}.
Also, as $\alpha$ increases, the powers under pure data decrease very little for all three hypotheses. 
The changes at small $\alpha>0$ under pure data with respect to the classical Wald test at $\alpha=0$ 
are clearly not quite significant. 
On the other hand, under contaminated data, the {levels} and powers of the classical Wald-test ($\alpha=0$)
change drastically for all hypotheses. But those for the proposed Wald-type tests at small positive $\alpha$ 
remain more stable under {both types of contaminations}.

\begin{table}[h]
	\caption{Empirical {levels} and powers for the MDPDE based Wald-type tests for different null hypotheses 
		and different $\alpha$ under {(mild) contamination scheme (I)}}
	\centering	
	{	
		\resizebox{\textwidth}{!}{
			\begin{tabular}{l|rrr|rrr|rrr|rrr}\hline
				&	\multicolumn{6}{c|}{$n=50$}	&	\multicolumn{6}{c}{$n=100$}\\\hline
				&	\multicolumn{3}{c|}{Size}	&	\multicolumn{3}{c|}{Power} &	\multicolumn{3}{c|}{Size}	&	\multicolumn{3}{c}{Power}\\
				$\alpha$	&	$H_0^{L1}$	&	$H_0^{L2}$	& $H_0^{L3}$	&	$H_0^{P1}$	&	$H_0^{P2}$	&	$H_0^{P3}$ 
				&	$H_0^{L1}$	&	$H_0^{L2}$	& $H_0^{L3}$	&	$H_0^{P1}$	&	$H_0^{P2}$	&	$H_0^{P3}$ \\\hline
				0 (Wald)	&	0.172	&	0.100	&	0.163	&	0.784	&	0.416	&	0.831	&	0.264	&	0.117	&	0.237	&	0.985	&	0.750	&	0.993	\\
				0.1	&	0.155	&	0.098	&	0.153	&	0.798	&	0.428	&	0.844	&	0.233	&	0.103	&	0.209	&	0.993	&	0.784	&	0.996	\\
				0.2	&	0.144	&	0.092	&	0.146	&	0.806	&	0.444	&	0.859	&	0.207	&	0.091	&	0.187	&	0.993	&	0.804	&	0.996	\\
				0.3	&	0.140	&	0.082	&	0.137	&	0.819	&	0.454	&	0.863	&	0.182	&	0.087	&	0.175	&	0.994	&	0.815	&	0.996	\\
				0.4	&	0.137	&	0.078	&	0.129	&	0.824	&	0.462	&	0.859	&	0.164	&	0.078	&	0.160	&	0.996	&	0.816	&	0.996	\\
				0.5	&	0.133	&	0.078	&	0.124	&	0.828	&	0.461	&	0.860	&	0.156	&	0.076	&	0.151	&	0.996	&	0.818	&	0.997	\\
				0.6	&	0.125	&	0.079	&	0.120	&	0.828	&	0.460	&	0.856	&	0.147	&	0.075	&	0.148	&	0.996	&	0.817	&	0.997	\\
				0.7	&	0.120	&	0.077	&	0.120	&	0.821	&	0.453	&	0.854	&	0.134	&	0.073	&	0.136	&	0.994	&	0.809	&	0.997	\\
				\hline
	\end{tabular}}}
	\label{TAB:testcont_data}
\end{table}

\begin{table}[h]
	
	\caption{Empirical {levels} and powers for the MDPDE based Wald-type tests for different null hypotheses 
		and different $\alpha$ under (extreme) contamination scheme (II)}
	\centering	
	\resizebox{\textwidth}{!}{
	\begin{tabular}{l|rrr|rrr|rrr|rrr}\hline
		&	\multicolumn{6}{c|}{$n=50$}	&	\multicolumn{6}{c}{$n=100$}\\\hline
		&	\multicolumn{3}{c|}{Size}	&	\multicolumn{3}{c|}{Power} &	\multicolumn{3}{c|}{Size}	&	\multicolumn{3}{c}{Power}\\
		$\alpha$	&	$H_0^{L1}$	&	$H_0^{L2}$	& $H_0^{L3}$	&	$H_0^{P1}$	&	$H_0^{P2}$	&	$H_0^{P3}$ 
		&	$H_0^{L1}$	&	$H_0^{L2}$	& $H_0^{L3}$	&	$H_0^{P1}$	&	$H_0^{P2}$	&	$H_0^{P3}$ \\\hline
		0 (Wald)	&	0.992	&	0.945	&	0.977	&	0.025	&	0.199	&	0.240	&	0.998	&	0.996	&	0.998	&	0.009	&	0.133	&	0.371	\\
		0.1	&	0.983	&	0.889	&	0.938	&	0.022	&	0.165	&	0.192	&	0.996	&	0.971	&	0.991	&	0.023	&	0.078	&	0.376	\\
		0.2	&	0.836	&	0.719	&	0.767	&	0.088	&	0.180	&	0.235	&	0.779	&	0.699	&	0.739	&	0.268	&	0.135	&	0.570	\\
		0.3	&	0.453	&	0.401	&	0.426	&	0.483	&	0.371	&	0.561	&	0.355	&	0.279	&	0.295	&	0.761	&	0.552	&	0.875	\\
		0.4	&	0.238	&	0.220	&	0.236	&	0.754	&	0.462	&	0.786	&	0.157	&	0.133	&	0.132	&	0.931	&	0.739	&	0.969	\\
		0.5	&	0.151	&	0.141	&	0.150	&	0.852	&	0.472	&	0.901	&	0.124	&	0.092	&	0.091	&	0.978	&	0.778	&	0.992	\\
		0.6	&	0.120	&	0.103	&	0.123	&	0.918	&	0.627	&	0.934	&	0.103	&	0.074	&	0.080	&	0.989	&	0.785	&	0.996	\\
		0.7	&	0.100	&	0.094	&	0.105	&	0.916	&	0.608	&	0.938	&	0.095	&	0.066	&	0.081	&	0.993	&	0.784	&	0.997	\\
		\hline
	\end{tabular}}
	\label{TAB:testcont_data2}
\end{table}
%

{
\section{\textcolor{black}{Applications  to Real-life Data}}
\label{SEC:Data}
\subsection{Application \textcolor{black}{ 1: AIS Data {(The motivating Example)}}}
}

\begin{figure}[!h]
	\centering
	\includegraphics[width=0.32\linewidth]{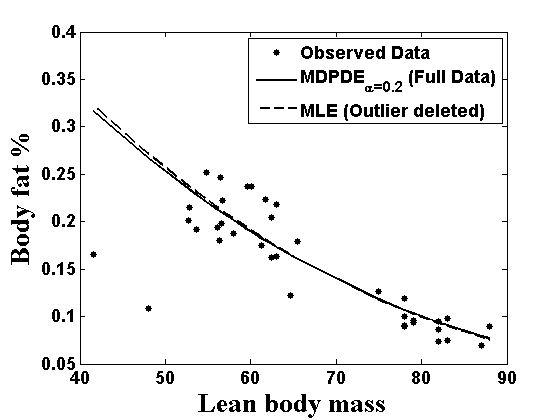}
	\includegraphics[width=0.32\linewidth]{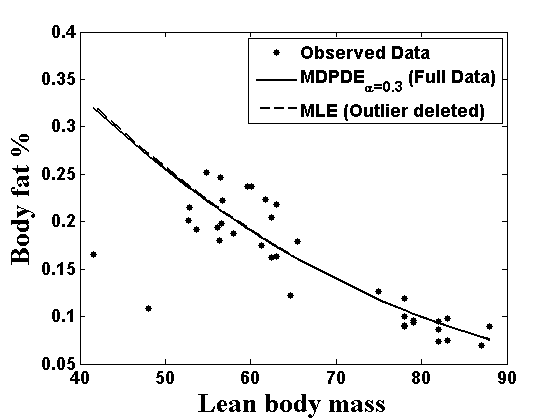}
	\includegraphics[width=0.32\linewidth]{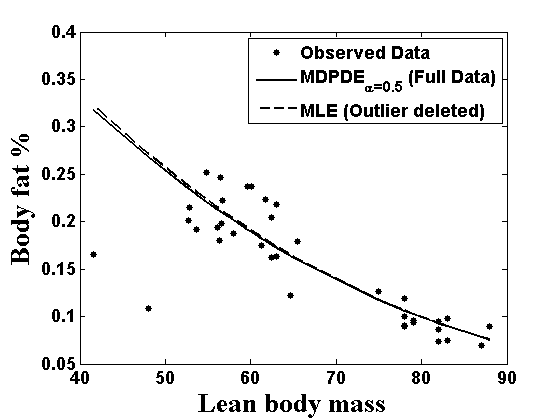}
	\caption{The BRM fitted lines for the AIS data based on the MDPDEs at $\alpha=0.2, 0.3, 0.5$ for the full data, 
	along with that based on the outlier deleted MLE.}
	\label{FIG:ais_mdpde}
\end{figure}

Let us start our illustration with reanalyzing the motivating AIS Dataset described in Section \ref{SEC:intro}.
We compute the proposed MDPDEs of the parameter $\boldsymbol{\theta}=(\beta_1, \beta_2, \phi)^T$ of 
the fitted BRM for different values of the tuning parameters $\alpha$ based on the full data and 
the outlier deleted data. The resulting estimates are reported in Table \ref{TAB:AIS_Est}
along with the {most commonly used} MLE (at $\alpha=0$).
Clearly, unlike the MLE, the proposed MDPDEs with $\alpha \geq 0.3$ change very little 
in the presence of two outlying observations. Further, the MDPDEs obtained based on the full data
are themselves very close to the outlier deleted MLE 
(See Figure \ref{FIG:ais_mdpde}) and so they can be used safely without bothering about the outliers.

\begin{table}[h]
	\caption{MDPDEs of $(\beta_1, \beta_2, \phi)^T$ for the AIS data, 
		along with the p-values for testing $H_0 : \beta_1=0$ using the Wald-type tests}
\centering	
\begin{tabular}{l|rrrr|rrrr}\hline
	&	\multicolumn{4}{c|}{Full Data}	&	\multicolumn{4}{c}{Outlier deleted data}\\
$\alpha$	&	$\beta_1$	&	$\beta_2$	& $\phi$	&p-value &	$\beta_1$	&	$\beta_2$	&	$\phi$ & p-value \\\hline
0 (MLE)	&	0.098	&	-0.027	&	96.616	& 0.699	&	0.838	&	-0.038	&	246.305	& 0 	\\
0.1	&	0.328	&	-0.031	&	116.026		& 0.158	& 0.832	&	-0.038	&	238.036	& 0 	\\
0.2	&	0.765	&	-0.037	&	206.180		& 0 	&	0.824	&	-0.038	&	231.658	& 0 	\\
0.3	&	0.807	&	-0.038	&	219.286		& 0 	&	0.815	&	-0.038	&	227.072	& 0 	\\
0.4	&	0.804	&	-0.038	&	218.032		& 0 	&	0.804	&	-0.038	&	224.270	& 0 	\\
0.5	&	0.794	&	-0.038	&	216.333		& 0 	&	0.790	&	-0.038	&	223.383	& 0 	\\
		\hline
	\end{tabular}
	\label{TAB:AIS_Est}
\end{table}

Next, let us consider the problem of testing significance of the intercept term, namely  $H_0 : \beta_1=0$.
{The p-value of the the existing MLE based Wald test} changes drastically due to the presence of two outliers.
We apply our proposed Wald-type tests based on the MDPDEs for this testing problem and 
resulting p-values are reported in Table \ref{TAB:AIS_Est}. Again, the proposed tests with $\alpha \geq 0.2$
generate stable p-values (which is zero) even in the full data with outliers. 
Therefore, the use of the proposed MDPDE and corresponding Wald-type tests with slightly larger $\alpha> 0$ 
can successfully tackle the effect of two outliers in the dataset yielding robust estimators and inference
even without separately finding and removing these outliers.

\begin{figure}[h!]
	\centering
	\subfloat[Clean Data]{
		\includegraphics[width=0.4\textwidth]{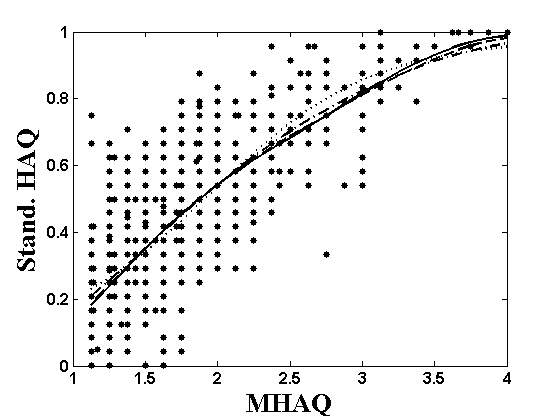}
		\label{FIG:HAQ_clean}}
	~ 
	\subfloat[Data with Outliers (red)]{
		\includegraphics[width=0.4\textwidth]{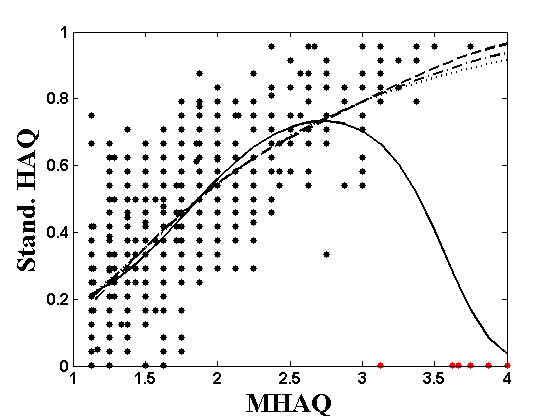}
		\label{FIG:HAQ_out}}
	\caption{The BRM fitted lines for the HAQ data based on the MDPDEs with different $\alpha$
		[Solid line: $\alpha=0$ (MLE); Dashed line: $\alpha=0.1$; Dash-dotted line: $\alpha=0.4$; Dotted line: $\alpha=0.7$].}
	\label{FIG:HAQ}
\end{figure}

{
\subsection{\textcolor{black}{Application 2: HAQ Dataset}}
}
In this example, we consider data on {a} certain standardized health assessment {questionnaire} (HAQ) 
from the Division for Women and Children at the Oslo University Hospital at Ulleval, Oslo, Norway.
The data, obtained from Prof.~Nils L.~Hjort of University of Oslo through personal communication, 
contain the  original (elaborative) HAQ scores along with an easy-to-use modified version (MHAQ) for 1018 patients.
These data have been used by \cite{Claeskens/Hjort:2008} to predict the original HAQ score  
from the simpler MHAQ scores, after suitable standardization, through a beta regression model. 
They have argued that the most healthy 219 patients with MHAQ = 1 need to be treated separately,
but the remaining 799 patients' data can be modelled well by a polynomial BRM with covariates
$\boldsymbol{x}=(1, \mbox{MHAQ}, \mbox{MHAQ}^2, \mbox{MHAQ}^3)^T$ and the logit link function. 
The corresponding fitted line based on the MLE is plotted in  Figure \ref{FIG:HAQ_clean};
clearly there is no outlier in the data. 
Here, the response variable HAQ takes the values in $[0, 3]$ inclusive of the end-points
and, to get it within the open interval $(0, 1)$, we use the popular ad-hoc transformation $y=((\mbox{HAQ}/3).(n-1)+0.5)/n$,
where $n=799$ is the total sample size \citep{Smithson/Verkuilen:2006, Melo/etc:2015}.
Now, let us {compute the} MDPDEs for this clean dataset to illustrate the behavior of our proposal in pure data.
The resulting estimators in fact turn out to be very close to the MLE 
which can clearly be seen from the fitted lines in Figure \ref{FIG:HAQ_clean}.

Now, to illustrate the robustness aspect, let us change only 6 largest HAQ values to (1$-$HAQ) values
and again derive the MLE and the MDPDEs; the fitted lines are shown in Figure \ref{FIG:HAQ_out}
(the artificial outliers are marked as red points).
Note that, only due to these 6 outliers, which is about only 0.75\% of the total number of observations,
the MLE changes to a drastically different fit which clearly gives an erroneous inference. 
In fact the MLE based Wald test for testing {the} significance of {the} intercept term now gives the p-value of $0.64$ 
(implying non-significance) with these outliers, which was zero (significant) in the original clean data.
However, the MDPDE based fits remain very  stable for all $\alpha\geq 0.1$ even in
the presence of these outliers as seen from Figure \ref{FIG:HAQ_out}.
Also, the {corresponding} MDPDE based Wald-type {tests at $\alpha\geq 0.1$ yield} correct p-value of zero for
testing the significance of intercept term both in the clean data and with these outliers.
\\

{
\subsection{\textcolor{black}{Application 3: Stress-Anxiety Data (Psychology)}}\label{SEC:Stress-Anxiety Data}
}

\begin{figure}[!h]
	\centering
	\includegraphics[width=0.4\linewidth]{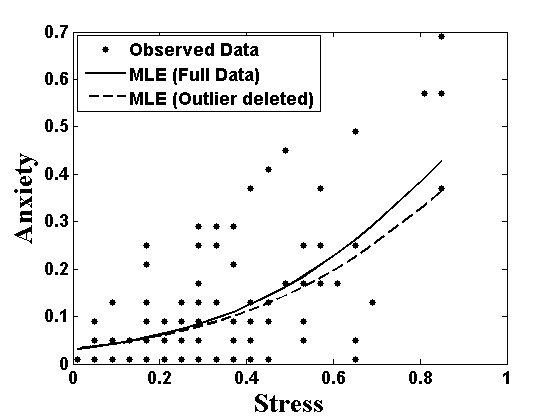}
	\includegraphics[width=0.4\linewidth]{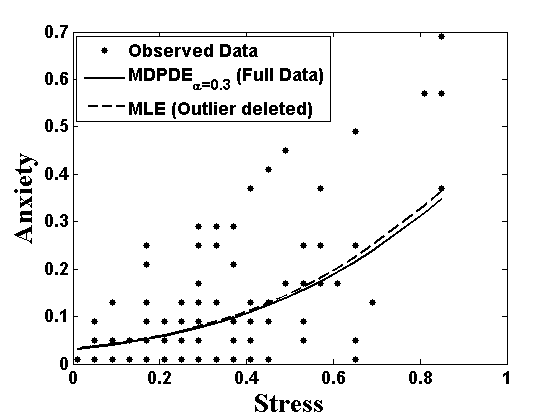}
	\caption{The BRM fitted lines for the Stress-Anxiety data based on the MDPDEs at $\alpha=0$ (MLE) and $\alpha=0.3$
		for the full data, along with that of the outlier deleted MLE.}
	\label{FIG:Stress_mdpde}
\end{figure}

Our final example is from a psychological trial among 166 nonclinical women in Australia
measuring the scores on suitable tests of their anxiety, depression and stress symptoms.
The details of the data can be found in \cite{Smithson/Verkuilen:2006}
who have analyzed it with a beta regression model with response as anxiety scores
and the covariates being the intercept and the stress scores along with the logit link function. 
\cite{Chien:2013} has studied these data to illustrate that there are several groups of highly 
influential observations affecting the MLE. We consider a set of 5 such outliers with higher anxiety scores 
and compute the MLE of the BRM parameters based on the full data and after deleting these outliers.
The corresponding fitted lines are shown in Figure \ref{FIG:Stress_mdpde}
which clearly {indicate} the non-robust nature of the MLE against the outlying observations.

We have applied our proposed MDPDE for these data and, as before, the MDPDEs with $\alpha>0.2$ {yield} robust estimators.
For brevity, we only present the fitted lines corresponding to the MDPDE with $\alpha=0.3$ based on the full data 
in Figure \ref{FIG:Stress_mdpde}; clearly the result is very close to that of the outlier deleted MLE
indicating the robustness of our proposal.

{
\subsection{\textcolor{black}{On the choice of the tuning parameter $\alpha$}}
}

The proposed DPD based robust estimators and Wald-type tests depend on a tuning parameter $\alpha$.
We have seen, both theoretically and empirically, 
that the efficiency of the proposed MDPDE under pure data decreases slightly as $\alpha $ increases,
but their robustness under contamination increases significantly.
Thus, the tuning parameter ${\alpha }$ yields a trade-off between efficiency and robustness of the proposed estimator. 
For hypothesis testing also, the asymptotic contiguous power decreases slightly with increasing $\alpha$, 
but the robustness of its level and power improves significantly under contamination;
here $\alpha$ trades off the contiguous power under pure data with robustness against outliers. 
Therefore, in either case, this tuning parameter $\alpha$ needs to be chosen appropriately for a given dataset.

As observed from various simulations and real data analyses,  
an $\alpha \approx 0.3,~0.4$ gives sufficiently robust estimator without significant loss in efficiency under pure data
and also provides a desired trade-off for the corresponding Wald-type test.
So, the empirical suggested value of ${\alpha }$ is to be taken around 0.3 to 0.4
which is expected to work well in most of the applications.

However, for a better trade-off based on the amount of contamination in the given dataset,
a data-driven choice of this tuning parameter $\alpha $ could be useful.
There are only a few such approaches for the DPD based inference.
We propose to follow the approach presented by \cite{Warwick/Jones:2005} and \cite{Ghosh/Basu:2015}
for the iid and the non-homogeneous data respectively. Their approach is mainly based on choosing $\alpha$ 
by minimizing an appropriate estimate of the MSE given by 
$$
E\left[(\widehat{\boldsymbol\theta}_{n,\alpha}-\boldsymbol\theta^*)^T
(\widehat{\boldsymbol\theta}_{n,\alpha}-\boldsymbol\theta^*)\right]
= (\boldsymbol{\theta}_{\alpha}-\boldsymbol\theta^*)^T(\boldsymbol{\theta}_{\alpha}-\boldsymbol\theta^*)
+\frac{1}{n} Trace\left[\boldsymbol\Psi_n^{-1}\boldsymbol\Omega_n\boldsymbol\Psi_n^{-1}\right],
$$
where $\boldsymbol\theta^*$ is the target parameter value, $\boldsymbol\theta_\alpha=\boldsymbol{T}_\alpha(G_a, \dots, G_n)$
and $\widehat{\boldsymbol\theta}_{n,\alpha}$ is the MDPDE with tuning parameter $\alpha$.
For the present case of beta regression models,
we can estimate this MSE by plugging in the MDPDE $\widehat{\boldsymbol\theta}_{n,\alpha}$ 
for $\boldsymbol\theta_\alpha$ and also in the variance part, 
but need to use different pilot estimators for ${\theta }^*$.
\cite{Ghosh/Basu:2015} have suggested that the use of the MDPDE with $\alpha=0.5$ serves well as the pilot estimator
in case of {the} linear regression model. This suggestion may be followed in the present case of BRM also,
but it needs substantial further investigation which we hope to do in our future research. 
\\


{
	\section{Extension to Non-Linear Variable Dispersion Beta Regressions}
\label{SEC:VarBRM}

Although till now we have restricted ourselves to the fixed dispersion (or precision) linear BRM (\ref{EQ:BRM}) for simplicity,
our proposed methodology is in no way limited to such restrictions and can easily be extended to various more complex BRMs.
Thus, it is indeed possible to fully exploit the flexibility of the beta regression models through such extensions. 
To illustrate this claim, in this section, we present the extension of the proposed MDPDE for 
a general class of non-linear and variable dispersion BRMs from \cite{Simas/etc:2010}.
For this class of BRMs, we allow the precision parameter $\phi$ (and hence also the dispersion) 
to be variable for different $y_i$ so that now we assume $y_i\sim f(y_i; \mu_i, \phi_i)$
and model $\phi_i$ by possibly another set of covariates, say $\boldsymbol{z}_i \in \mathbb{R}^q$,
through suitable link function $h$ (may be different from the link function $g$ used in the mean model).
We can also avoid the linearity constraint on the  predictors to have a larger flexible class of BRMs given by
\begin{eqnarray}
y_i \sim f\left(y_i; \mu_i, \phi_i\right) \mbox{ independently, with } 
{g}\left({{\mu }}_{{i}}\right){=} \eta_1({\boldsymbol{x}}_{i}, \boldsymbol{\beta }),
~{h}\left(\phi_i\right){=} \eta_2({\boldsymbol{z}}_{i}, \boldsymbol{\gamma }),
~~i=1, \ldots, n,
\label{EQ:BRMGen}
\end{eqnarray}	 
where $\eta_1$ and $\eta_2$ are some known functions 
and $\boldsymbol\beta \in \mathbb{R}^p$ and $\boldsymbol{\gamma}\in\mathbb{R}^q$ 
are the vectors of unknown regression coefficients corresponding to the mean and the precision models respectively.
Note that, the covariates can have different dimensions compared to the corresponding regression coefficients
(although we have kept them the same without any loss of generality), 
but we need to assume that the derivative matrices of $\eta_1$ and $\eta_2$ with respect to 
$\boldsymbol{\beta}$ and $\boldsymbol{\gamma}$, respectively, have ranks $p$ and $q$. 
Then, our parameter of interest becomes $\boldsymbol\theta =(\boldsymbol\beta^T, \boldsymbol{\gamma}^T )^T\in\mathbb{R}^{p+q}$.

Note that, under the general class of BRMs (\ref{EQ:BRMGen}) also, assuming the covariates to be fixed,
the observed responses $y_i$s are independent but non-homogeneous with model density of $y_i$ being 
$f_i(\cdot, \boldsymbol{\theta}) \equiv  f(\cdot; \mu_i, \phi_i)$ for $i=1, \ldots, n$.
So this again belongs to the general set-up of \cite{Ghosh/Basu:2013} and, as before, 
we can define the MDPDE of $\boldsymbol{\theta}$ by minimizing the average DPD measure 
between the $i$-th data point and the corresponding model density $f_i(\cdot, \boldsymbol{\theta})$ for $i=1, \ldots, n$.
Following the general theory of \cite{Ghosh/Basu:2013}, as presented in Section 1 of the online supplement,
the MDPDE objective function  under the BRMs (\ref{EQ:BRMGen}) can again be simplified to have the form
\begin{eqnarray}
H_{n,\alpha}(\boldsymbol\theta) = n^{-1}\sum^n_{i=1}
\left[\widetilde{K_{i,\alpha}}(\boldsymbol\theta) - \frac{1+\alpha}{\alpha}f_i(y_i, \boldsymbol\theta)^\alpha \right],
\label{EQ:HnG}
\end{eqnarray}
where now we have   $f_i(\cdot, \boldsymbol{\theta}) \equiv  f(\cdot; \mu_i, \phi_i)$
and $\widetilde{K_{i,\alpha }}(\boldsymbol\theta)
=\frac{B\left((1+\alpha)\mu_i\phi_i -\alpha , (1+\alpha)\left(1-\mu_i\right)\phi_i -\alpha \right)}{
	B{\left({\mu }_i\phi_i , \left(1- \mu_i\right)\phi_i \right)}^{\alpha}}.$
The corresponding estimating equations for the BRMs (\ref{EQ:BRMGen}), 
obtained from the general Equation (2) of the online supplement,
are again  given by
\begin{eqnarray}
\sum^{n}_{i=1}\left[\widetilde{\gamma^{(\alpha)}_{1,i}}(\boldsymbol\theta) - \left(y^*_{1,i} - \widetilde{\mu^*_{1,i}}\right)
\frac{\phi_i}{g'(\mu_i)}f_i(y_i, \boldsymbol\theta)^\alpha\right]
\frac{\partial\eta_1({\boldsymbol{x}}_{i}, \boldsymbol{\beta })}{\partial\boldsymbol{\beta}} 
&=& {\boldsymbol{0}_p},\label{EQ:Est_Eqn1G}\\
\sum^n_{i=1}\left[\widetilde{\gamma^{(\alpha)}_{2,i}}(\boldsymbol\theta) 
- \left\{\mu_i\left(y^*_{1,i} - \widetilde{\mu^*_{1,i}}\right)+ \left(y^*_{2,i} - \widetilde{\mu^*_{2,i}}\right)
\right\}\frac{1}{h'(\phi_i)}f_i(y_i, \boldsymbol\theta)^\alpha\right] 
\frac{\partial\eta_1({\boldsymbol{x}}_{i}, \boldsymbol{\gamma})}{\partial\boldsymbol{\gamma}} 
&=& \boldsymbol{0}_q, \label{EQ:Est_Eqn2G}
\end{eqnarray}
where we now have 
$\widetilde{{\mu }^*_{1,i}}=E\left(y^*_{1,i}\right)
=\psi \left({\mu }_i\phi_i \right) - \psi \left(\left(1-{\mu }_i\right)\phi_i \right)$,
$\widetilde{{\mu }^*_{2,i}}=E\left(y^*_{2,i}\right)
=\psi \left(\left(1-{\mu }_i\right)\phi_i \right) - \psi \left(\phi_i \right)$,
$\widetilde{\gamma^{\left(\alpha \right)}_{1,i}}(\boldsymbol\theta) 
= \left(\psi\left(\widetilde{a_{i,\alpha}}\right) - \psi\left(\widetilde{b_{i,\alpha}}\right) - \widetilde{\mu^*_{1,i}}\right)
\frac{\phi_i \widetilde{K_{i,\alpha}}(\boldsymbol\theta )}{g'(\mu_i)}$
and
\begin{eqnarray}
\widetilde{{\gamma }^{\left(\alpha \right)}_{2,i}}\left(\boldsymbol\theta \right) 
&=& \left[ {\mu }_i\left(\psi\left(\widetilde{a_{i,\alpha }}\right)-\psi \left(\widetilde{b_{i,\alpha }}\right) 
- \widetilde{\mu^*_{1,i}}\right) + \left(\psi\left(\widetilde{b_{i,\alpha }}\right) 
-\psi\left(\widetilde{a_{i,\alpha }} + \widetilde{b_{i,\alpha }}\right) - \widetilde{\mu^*_{2,i}}\right)\right]
\frac{\widetilde{K_{i,\alpha}}(\boldsymbol\theta )}{h'(\phi_i)}
\nonumber
\end{eqnarray}
with $\widetilde{a_{i,\alpha }}=\left(1+\alpha \right)\mu_i\phi_i -\alpha$, 
$\widetilde{b_{i,\alpha}}=(1+\alpha)(1-\mu_i)\phi_i -\alpha$.
Proceeding similarly, we can derive all asymptotic and robustness properties of the MDPDEs
under the general class of flexible BRMs (\ref{EQ:BRMGen}), as before, using the general results
from the online supplement.
Suitable robust Wald-type tests of any hypothesis under BRMs (\ref{EQ:BRMGen})
can also be developed with similar properties based on the general results from Section 2 of the online supplement.
Considering the length of the current manuscript, we have decided to keep their details for a future report;
but the general interpretations and developments are expected to be exactly similar
(as observed in the following example).

\bigskip\noindent
\textbf{Example: Stress-Anxiety Data with Variable-Dispersion Beta Regression Model}\\
As an illustration of the performance of the MDPDEs under the variable dispersion BMRs,
let us reconsider the Stress-Anxiety data studied in Section \ref{SEC:Stress-Anxiety Data}.
\cite{Smithson/Verkuilen:2006} have shown that the anxiety scores in this data set 
can be modeled better with a (linear) variable dispersion beta-regression model than the fixed dispersion BRM 
(as done in Section \ref{SEC:Stress-Anxiety Data}); 
this is because the variability in anxiety scores clearly depends on the level of stress-scores 
(see Figure \ref{FIG:Stress_mdpde}).
So, we now fit the general model (\ref{EQ:BRMGen}) for this dataset
with $y_i=\mbox{Anxiety-Score}_i$, $\boldsymbol{x}_i =\boldsymbol{z}_i= (1, \mbox{Stress-Score}_i)^T$ 
and $g$ and $h$ being the `logit' and `log' link functions respectively.
Then, we compute  the MDPDEs at different $\alpha>0$ and the MLEs (at $\alpha=0$) of the regression parameters 
$\boldsymbol{\theta}=(\beta_1, \beta_2, \gamma_1, \gamma_2)^T$ under the full data and the outlier deleted data.
Since the changes in the estimators are small, in order to illustrate the extent of robustness,
we here study the relative change in the estimators under full data with that under outlier deleted data,
which is presented in Table \ref{TAB:StressAnx_VarDisp} for $\alpha=0, 0.3$.
Clearly the change due to outliers is significantly reduced for all parameters,
specially for both the slope parameters, while using the newly proposed MDPDEs with $\alpha\approx 0.3$.
All estimates are statistically significant indicating suitability of the fitted model.

%

}

\begin{table}[h]
	
	\caption{Relative difference due to outliers, in the MLE and the MDPDE at $\alpha=0$, for the Stress-Anxiety data 
		under the variable-dispersion BRM}
	\centering	
	\begin{tabular}{l|cccc}\hline
						&	$\beta_1$	&	$\beta_2$	&	$\gamma_1$	&	$\gamma_2$\\\hline
	MLE						&	1.23\%		&	1.89\%		&	4.47\%		&	13.65\%		\\
	MDPDE$_{\alpha=0.3}$	&	0.92\%		&	0.29\%		&	3.41\%		&	6.74\%		\\\hline
	\end{tabular}
	\label{TAB:StressAnx_VarDisp}
\end{table}

{
	\section{\textcolor{black}{Concluding remarks}}
\label{SEC:concluding}
}

In this paper, we have developed a robust statistical inference procedure under the beta regression model
for modeling responses on $(0, 1)$. We have proposed the minimum DPD estimator for estimating the parameters in the {fixed dispersion} BRM
and developed a class of Wald-type tests based on them for testing general linear hypotheses in regression coefficients.
Beside discussing their asymptotic properties, we have also justified the robustness of the proposed methodology through
appropriate influence function analyses. Suitable numerical illustrations have been provided 
along with three important real data applications from health-care studies.
{Some indications are also provided, with application, for extending the proposed inference to 
	the variable dispersion beta regression models having non-homogeneous precisions.}

{
	
It is worthwhile to note that an important measure of global robustness of an inference procedure
is their breakdown point, which is not explored in this paper.
\cite{Ghosh/Basu:2013}	have shown that the proposed DPD based inference with $\alpha>0$
has the maximum possible breakdown point of $0.5$ under mild boundedness conditions on the covariates 
in a fixed-design linear regression model.
We hope that similar breakdown result can also be derived for the BRM under certain conditions (a mathematical challenge),
but we do not have an explicit proof at this moment.

However, our proposed methodology can be directly applied to any complex big dataset 
to generate robust inference without bothering much about outliers in the data.
This is because the proposed estimator has a simple unbiased estimating equation
which can be easily solved efficiently for such big datasets using appropriate numerical techniques
and the underline objective function also helps us to avoid any problem in cases
with multiple roots to this estimating equation.
But, for high dimensional datasets with more covariates than observations, 
we need to add a suitable regularization penalty factor (like LASSO or SCAD penalties)
in the proposed objective function (\ref{EQ:Hn}).
Such penalized DPD based approach for robust inference under high-dimensional linear regression model
has recently been studied by \cite{Zang/etc:2017}.
Similar extension under the present BRM with high-dimensional structure will be an interesting future work.
	
	}

{Besides detailed study of the extension discussed in Section \ref{SEC:VarBRM}, }
it will {also} be {very} useful to further extend {it}
to develop robust inference for \trd{more general beta regression models, like}
the inflated zero or one (or both) BRMs for datasets containing 0 or 1 or both values
{and the BRMs with repeated measurements; the general theory presented in the online supplement 
will directly guide in these extensions.}
\trd{or the variable dispersion beta regression models for non-homogeneous precision parameter $\phi$ etc.}
Also, the proposed scheme for selection of a data-driven choice of the tuning parameter $\alpha$
needs more investigation. We plan to pursue some of these extensions in our future works.


\bigskip\noindent
\textbf{Acknowledgment:} 
The author wants to express his sincere thanks to Prof.~Nils L.~Hjort of University of Oslo for the HAQ dataset and 
Prof.~Ayanendranath Basu of Indian Statistical Institute for several constructive suggestions and comments about the work.
The author also wishes to thank the Editor and three anonymous referees for their careful reading of 
the manuscript and several constructive suggestions which have significantly improved the paper. 

\bigskip\noindent
\textbf{Funding:} 
This work is supported by the INSPIRE Faculty research grant from the Department of Science and Technology, Govt. of India.

\end{document}